\documentclass[iop]{emulateapj}

\newcommand{\kms}{km~s$^{-1}$}
\newcommand{\msunyr}{$M_\sun$~yr$^{-1}$}
\newcommand{\nod}{\nodata}

\slugcomment{Accepted by ApJ, 2011 January 21}

\begin{document}

\title{Characterizing the {\em IYJ} Excess Continuum Emission in T Tauri Stars}
\author{William Fischer\altaffilmark{1,2,3}, Suzan Edwards\altaffilmark{2,3,4}, Lynne Hillenbrand\altaffilmark{3,5}, John Kwan\altaffilmark{6}}
\altaffiltext{1}{Dept.\ of Physics and Astronomy, University of Toledo, Toledo, OH 43606, wfische@utnet.utoledo.edu}
\altaffiltext{2}{Visiting Astronomer, NASA Infrared Telescope Facility}
\altaffiltext{3}{Visiting Astronomer, Keck Observatory}
\altaffiltext{4}{Five College Astronomy Dept., Smith College, Northampton, MA 01063, sedwards@smith.edu}
\altaffiltext{5}{Dept.\ of Astronomy, California Institute of Technology, Pasadena, CA 91125, lah@astro.caltech.edu}
\altaffiltext{6}{Five College Astronomy Dept., University of Massachusetts, Amherst, MA 01003}
\addtocounter{footnote}{6}

\begin{abstract}
We present the first characterization of the excess continuum emission of accreting T Tauri stars between optical and near-infrared wavelengths.  With nearly simultaneous spectra from 0.48 to 2.4~\micron\ acquired with HIRES and NIRSPEC on Keck and SpeX on the IRTF, we find significant excess continuum emission throughout this region, including the $I$, $Y$, and $J$ bands, which are usually thought to diagnose primarily photospheric emission. The $IYJ$ excess correlates with the excess in the $V$ band, attributed to accretion shocks in the photosphere, and the excess in the $K$ band, attributed to dust in the inner disk near the dust sublimation radius, but it is too large to be an extension of the excess from these sources.  The spectrum of the excess emission is broad and featureless, suggestive of blackbody radiation with a temperature between 2200 and 5000 K.  The luminosity of the $IYJ$ excess is comparable to the accretion luminosity inferred from modeling the blue and ultraviolet excess emission and may require reassessment of disk accretion rates.  The source of the $IYJ$ excess is unclear. In stars of low accretion rate, the size of the emitting region is consistent with cooler material surrounding small hot accretion spots in the photosphere. However, for stars with high accretion rates, the projected area is comparable to or exceeds that of the stellar surface. We suggest that at least some of the $IYJ$ excess emission arises in the dust-free gas inside the dust sublimation radius in the disk.
\end{abstract}


\section{INTRODUCTION}

Classical T Tauri stars (CTTS) are low-mass stars in the final stages of disk accretion.  Accretion from the disk to the star is thought to be governed by the stellar magnetosphere, where a sufficiently strong magnetic field truncates the disk at several stellar radii and guides infalling material along field lines to the stellar surface at high latitudes, terminating in accretion shocks. Historically, a major factor contributing to the understanding that T Tauri stars are accreting matter from their circumstellar disks was the spectrum of their excess continuum emission from ultraviolet to radio wavelengths \citep{ber89}.  While the disks themselves are the primary source of continuum radiation at wavelengths longward of 2~\micron, optical and ultraviolet continuum emission in excess of the photosphere is attributed to accretion shocks on the stellar surface. Spectral lines also played a key role in developing the current magnetospheric accretion paradigm, providing the kinematic evidence that the stellar magnetosphere channels accreting material from the disk to the star and that accretion-powered winds arise from near-stellar regions \citep{naj00}. In this paper we focus on the excess continuum emission in a region that has not received close scrutiny to date -- the spectral region between the optical and the near infrared, where the photospheric emission from these late-type stars peaks. 

The well studied optical and ultraviolet excesses are used to derive a fundamental parameter of T Tauri stars -- the rate of accretion from the disk to the star. The inferred accretion rates span several orders of magnitude, from extremes of $10^{-10}$ to $10^{-7}$ \msunyr, with a median of $\sim 10^{-8}$ \msunyr\ at 1 Myr that declines with increasing age \citep{har98}. The most reliable estimates for disk accretion rates come from comparisons of spectrophotometric observations over a broad wavelength range to shock models. This was done successfully by Calvet and Gullbring (1998, hereafter \citealt{cal98}), who accounted for the excess continuum between 0.32 and 0.52~\micron\ with shock models that attribute the Paschen continuuum to optically thick post-shock gas in the heated photosphere and the Balmer continuum and Balmer jump to optically thin gas in the pre-shock and attenuated post-shock regions.  This combination of spectrophotometry with shock models gives a fairly uniform result for the spectrum of the excess in the Paschen continuum, with temperatures $\sim6000-8000$ K and a shape that is essentially blackbody. A more commonly used approach to determine accretion rates is to evaluate the excess emission over a limited range of wavelength in the Paschen continuum by comparing the depth of photospheric absorption lines to those of a template matched in temperature, gravity, and projected rotational velocity.  In the presence of a continuum excess, photospheric features will be weakened, and a quantity known as {\em veiling}, defined as the flux ratio $r=F_{\rm excess}/F_*$, can be derived \citep{bas90,har91}.  Accretion rates are then determined from a bolometric correction based on an isothermal slab model at an assumed temperature, density, and optical depth (\citealt{val93,har95}; Gullbring et al.\ 1998a, hereafter \citealt{gul98a,har03,her08}).  

At near-infrared wavelengths, CTTS show excess emission from 2 to 5 \micron\ that is well described by $\sim1400$~K blackbody radiation and is attributed to a raised rim of dust at the dust sublimation radius in the inner disk (Muzerolle et al.\ 2003, hereafter \citealt{muz03,fol01,joh01}). The magnitude of the excess is proportional to the accretion luminosity, requiring that the dust be heated by radiation from both the photosphere and the accretion shock.  $K$-band interferometry, which locates dust in a ring at a few tenths of an AU from the star \citep{eis05, mil07}, provides further evidence that 2 to 5 \micron\ emission in CTTS comes from sublimating dust in the inner disk. 

Evidence has been accumulating that there is an unexplained source of excess emission in CTTS between the optical and near infrared that cannot be attributed to shocks at the base of magnetic funnel flows or dust sublimating in the disk. As early as 1990, \citeauthor{bas90} found that the veiling longward of 0.5 \micron\ does not steadily decline with wavelength as expected from hot accretion shock models but instead is relatively constant between 0.5 and 0.8 \micron, and high veiling at 0.80--0.85 \micron\ has more recently been reported by \citet{har03} and \citet{whi04}. Furthermore, \citet{edw06} found high veiling at 1~\micron, which is unlikely to come from 1400 K dust in the inner disk since this emission will fall rapidly to short wavelengths from its peak around 3~\micron. Also, $K$-band interferometric studies of T Tauri stars find the angular size of the near-infrared emission in modeling of multiple-baseline observations \citep{ake05} and the size-wavelength behavior in spectrally dispersed observations \citep{eis09} to suggest that gaseous material inside the dust sublimation radius is present at a temperature higher than that of the dust.  However, the lack of a systematic study of the wavelength dependence of excess emission between 0.5 and 2 \micron\ has inhibited the understanding of its role in accreting systems.  

The presence of unexplained continuum emission in CTTS between 0.5 and 2 \micron\ not only offers the opportunity to improve our understanding of the structure of accretion disk systems; it also poses a practical problem.  If we do not understand the shape of the veiling spectrum at wavelengths longer than 0.5 \micron, then we cannot reliably convert veiling measurements at these wavelengths via a simple bolometric correction into accurate accretion luminosities and disk accretion rates. There may also be ramifications for extinction determinations of CTTS if they are based on far-red colors assumed to be predominantly photospheric. 

In this paper we determine the excess spectrum over a broad wavelength region between 0.48 and 2.4 \micron\ for a sample of 16 classical T Tauri stars in Taurus.  The presentation includes \S\ 2 describing the sample and data reduction, \S\ 3 describing the derivation of the veiling and the excess emission spectra, \S\ 4 characterizing the behavior of the excess emission spectra, and \S\ 5 describing simple models for the excess, followed by a discussion in \S\ 6 and conclusions in \S\ 7. We demonstrate that the two traditional sources of optical and near-infrared excess emission, a hot component arising from an accretion shock-heated photosphere with small filling factor and a cool component arising from the dust sublimation radius of the accretion disk, are not sufficient to describe the observed excess. A third component of intermediate temperature seems to be required to fit the excess emission in this region, with a luminosity of the same order of magnitude as derived from the hot shock-heated gas.

\section{SAMPLE AND DATA REDUCTION}

\begin{deluxetable*}{lcccccc}
\tablecaption{SpeX CTTS Sample\label{t.sample}}
\tabletypesize{\footnotesize}
\tablewidth{0pt}
\tablehead{\colhead{Object} & \colhead{Spectral Type} & \colhead{$\log \dot{M}_{\rm{acc}}$} & \colhead{Refs.} & \colhead{SpeX} & \colhead{HIRES} & \colhead{NIRSPEC} \\ \colhead{(1)} & \colhead{(2)} & \colhead{(3)} & \colhead{(4)} & \colhead{(5)} & \colhead{(6)} & \colhead{(7)}}
\startdata
AA Tau\dotfill   & K7 & $-$8.5  & 5,2 & 27 & 30 & 30  \\
AS 353A\dotfill  & K5 & $-$5.4  & 3,3 & 26,27 & 30 & 30,01 \\
BM And\dotfill   & G8 & $>-$9 & 6,1 & 27 & 30 & 30 \\
BP Tau\dotfill   & K7 & $-$7.5  & 5,2 & 26 & 30 & 30 \\
CW Tau\dotfill   & K3 & $-$6.0  & 5,3 & 26 & 30,01 & 30,01 \\
CY Tau\dotfill   & M1 & $-$8.1  & 5,2 & 27 & 30 & 30 \\
DF Tau\tablenotemark{a}\dotfill   & M2 & $-$6.9  & 4,4 & 26 & 30 & 30 \\
DG Tau A\dotfill   & K7 & $-$5.7  & 5,3 & 26 & 30 & 30 \\
DK Tau A\dotfill   & K7 & $-$7.4  & 5,2 & 26 & 30 & 30 \\
DL Tau\dotfill   & K7 & $-$6.7  & 5,3 & 27 & 30 & 30 \\
DO Tau\dotfill   & M0 & $-$6.8  & 5,2 & 27 & 01 & 01 \\
DR Tau\dotfill   & K7 & $-$5.1  & 5,3 & 27 & 30 & 30 \\
HN Tau A\dotfill   & K5 & $-$8.6  & 7,7 & 27 & 30 & 30 \\
LkCa 8\dotfill   & M0 & $-$9.1  & 5,2 & 27 & 30 & 30 \\
RW Aur A\dotfill & K1 & $-$7.5  & 7,7 & 26 & \nod & \nod \\
UY Aur A\dotfill   & M0 & $-$7.6  & 4,4 & 27 & \nod & 01
\enddata
\tablecomments{Col.~2: Generally accurate to $\pm1$ subclass; Col.~3: Logarithm of the mass accretion rate in \msunyr; Col.~4: References for the spectral type and mass accretion rate; Cols.~5--7: UT day of the month on which a spectrum was obtained in 2006: Nov 26, 27, or 30 or Dec 01.} 
\tablenotetext{a}{Unresolved binary (0.09'' separation); properties are for the primary.}
\tablerefs{(1) \citealt{gue93}; (2) \citealt{gul98a}; (3) \citealt{har95}; (4) \citealt{har03}; (5) \citealt{ken95}; (6) \citealt{ros99}; (7) \citealt{whi01}.}
\end{deluxetable*}

Our sample consists of the 16 classical T Tauri stars in Table~\ref{t.sample}.  They were chosen to have a broad range of published mass accretion rates, covering four orders of magnitude.  All were observed with SpeX \citep{ray03} at the Infrared Telescope Facility (IRTF) on 2006 November 26 and 27.  A few days later, on November 30 and December 1, most were also observed nearly simultaneously with HIRES \citep{vog94} on Keck I and NIRSPEC \citep{mcl98} on Keck II. We combine the three data sets to address the excess emission over a broad wavelength range.

Seven of the objects in the sample are members of binary systems.  All but one of the systems have separations $\ge0.88\arcsec$ \citep{whi01}, and they were resolved by all three instruments.  In these cases we observed only the primary, rotating the spectrograph slit if necessary to avoid contamination by the secondary. They are noted with an ``A" in Table~\ref{t.sample}. DF Tau A and B, on the other hand, were unresolved at a separation of only 0.09\arcsec, and thus both components contribute to the spectra. The components have similar spectral types (M2.0 and M2.5; \citealt{har03}) and a flux ratio at $K$ of 1.62 \citep{whi01}.  In this work we treat DF Tau as a single star with the published properties of the primary.

\subsection{SpeX\label{s.spex}}

The SpeX data were taken in the short-wavelength cross-dispersed (SXD) mode, featuring a 0.3$\arcsec\times15\arcsec$ slit and yielding spectra that extend from 0.8 to 2.4~$\micron$ at a resolving power $R=2000$.  The detector is a $1024\times1024$-pixel Aladdin InSb array with a 0.15$\arcsec$ pixel scale. Total exposure times ranged from 16 to 48 minutes for the program stars, with  $8.2\leq J \leq10.7$, yielding a continuum $S/N > 250$ at $J$, increasing to longer wavelengths.

To facilitate the subtraction of sky emission lines, the observing sequence for a given object consisted of multiple 120 s exposures.  After the first and third exposures of a four-exposure sequence, the telescope was nodded by 8$\arcsec$ (an ABBA pattern) such that the object remained in the slit.  Normal A0 stars near the targets and at similar airmass (usually a difference of less than 0.05) were observed to allow the removal of atmospheric absorption lines and provide an approximate flux calibration.

The data were reduced with Spextool \citep{cus04}, an IDL package containing routines for dark subtraction and flat fielding, spectral extraction, wavelength calibration with arc lamp lines, and the combining of multiple observations from a nodding sequence.  The correction for atmospheric absorption lines was handled by {\em xtellcor} \citep{vac03}, an IDL routine that divides the target spectrum by a calibrator.  If the calibrator is an A0 star, then its spectrum consists to a good approximation only of hydrogen lines and atmospheric absorption lines.  Division by such a calibrator, if observed at approximately the same airmass as the target, removes the target's atmospheric absorption lines.  The hydrogen lines in the calibrator are removed by fitting a scaled model of Vega.  After telluric correction, the merging of the six orders provided in the SXD mode and the removal of remaining bad pixels are accomplished with separate programs in the Spextool package. None of the targets showed extended line emission in their 2D spectra.

The error in the flux calibration is dominated by different slit losses for the target and the calibrator.  We compared our approximate flux calibration to the $J$ and $K$ magnitudes from 2MASS\footnote{The Two Micron All Sky Survey (2MASS) is a joint project of the University of Massachusetts and the Infrared Processing and Analysis Center/California Institute of Technology, funded by the National Aeronautics and Space Administration and the National Science Foundation.}. Although we find that  $J_{\rm SpeX} - J_{\rm 2MASS}=-0.11\pm0.43$ and $K_{\rm SpeX} - K_{\rm 2MASS}=-0.06\pm0.41$, the discrepancy is much less between our data and 2MASS in the $J-K$ color, only $-0.05\pm0.20$ magnitudes, indicating that the magnitude differences are nearly color-independent, whether from variability or slit losses. Thus we are confident that despite uncertainties in the absolute flux calibration, the {\em shape} of the spectral energy distribution (SED) is preserved, and our continuum-normalized spectra fairly represent the wavelength dependence of the observed spectrum. 

\subsection{Echelle Spectra: HIRES + NIRSPEC\label{s.hires}}

Beginning three nights after the SpeX run, high-resolution red optical and $Y$-band echelle spectra of 14 of the 16 CTTS from the SpeX sample were obtained simultaneously with HIRES on Keck I and NIRSPEC on Keck II, respectively.  A 15th star from the sample was observed with NIRSPEC only. Stars with high-resolution spectra are noted in Table~\ref{t.sample}.

HIRES was used with the red collimator and the C5 decker ($1.15\arcsec\times7.0\arcsec$), which has a projected slit width of 4 pixels and a spectral resolution of $R=33,000$ ($\Delta V=8.8$~\kms). With the cross-disperser set to approximately $0.884^\circ$ and the echelle angle at $0.0^\circ$, we have nearly complete spectral coverage from about 0.48 to 0.92~\micron\ in 38 orders (17 on the blue chip, 12 on the green chip, and 7 on the red chip).  The red, green, and blue detectors were used in low-gain mode, resulting in readout noise levels of 2.8, 3.1, and 3.1 $e^{-1}$, respectively. Internal quartz lamps were used for flat fielding, and ThAr lamp spectra were used for wavelength calibration. The HIRES data were reduced with the MAuna Kea Echelle Extraction reduction script (MAKEE) written by Tom Barlow. Scott Dahm was the observer on Keck I during our HIRES run.

NIRSPEC was used with the N1 filter ($Y$ band), which covers the range 0.95 to 1.12~\micron\ in 14 spectral orders at a resolution $R=25,000$ ($\Delta V=12$~\kms).  Data reduction, including wavelength calibration and spatial rectification, extraction of one-dimensional spectra from the images, and removal of telluric emission and absorption features, is performed with the IDL package REDSPEC by S. S. Kim, L. Prato, and I. McLean, as discussed in \citet{fis08}.

\section{DERIVING THE EXCESS EMISSION SPECTRA\label{s.merge}}

We first determine the shape and intensity of the spectrum of excess (non-photospheric) continuum emission  between 0.48 and 2.4 \micron\ using both echelle and SpeX spectra.  We follow different procedures for the echelle and SpeX data and merge the results to give the SED of the excess emission over our full wavelength range.  The starting point for both data sets is to measure individual line veilings $r_\lambda$, defined as the ratio of excess flux $E_\lambda$ to photospheric flux $P_\lambda$ at a particular wavelength. Thereafter, we follow different routes to find the continuous veiling spectrum $V_\lambda$ (a function describing $r_\lambda$ over a continuous span of wavelength), and the spectrum of the excess continuum $E_\lambda$.  In both cases the excess continuum is expressed in units of the photospheric flux at 0.8~\micron.  Each step is described more fully in the following subsections.

\subsection{Echelle Spectra}

The determination of $E_\lambda$ from the echelle spectra follows directly from the veiling in each order where a reliable measurement can be made. To determine the echelle line veilings $r_\lambda$, we follow \citet{har91}, matching each echelle order of an accreting star to that of a non-accreting standard of comparable spectral type to which has been added a constant excess continuum that reproduces the depth of the photospheric features in the accreting star. For this procedure, we use a modified version of an interactive IDL routine kindly provided by Russel White to accomplish the requisite velocity shift of the standard via cross-correlation and the requisite rotational broadening of the standard before determining the required level of excess continuum.  A polynomial is then fit to the individual $r_\lambda$ with wavelength, yielding $V_\lambda$, and the excess emission $E_\lambda$ is the product of $V_\lambda$ and a temperature-matched photospheric template from the Pickles library \citep{pic98}. 

\begin{figure*}
\plotone{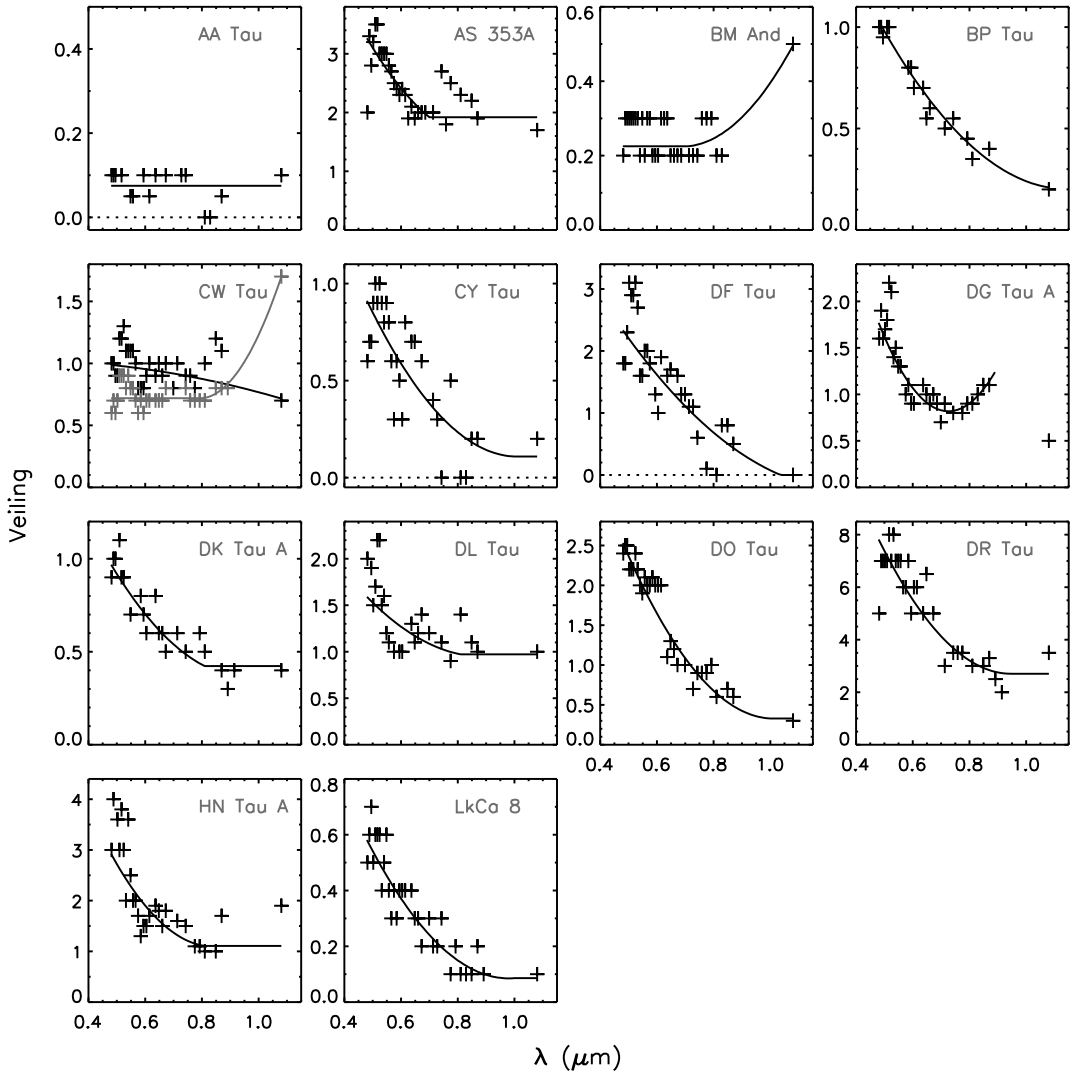}
\figcaption{Echelle veilings for the 14 CTTS observed with HIRES and NIRSPEC.  Measured values are shown with $+$ signs, and polynomial fits are shown with solid lines.  For CW Tau, results from a second echelle spectrum are overplotted in gray.\label{f.echelle}}
\end{figure*}

\begin{deluxetable*}{lccccccccccc}
\tablecaption{HIRES + NIRSPEC Average Line Veilings\label{t.hiresveil}}
\tabletypesize{\footnotesize}
\setlength{\tabcolsep}{0.075in}
\tablewidth{0pt}
\tablehead{\colhead{Object} & \colhead{UT Date} & \colhead{$r_{4883}$} & \colhead{$r_{5165}$} & \colhead{$r_{5485}$} & \colhead{$r_{5942}$} & \colhead{$r_{6365}$} & \colhead{$r_{6725}$} & \colhead{$r_{7427}$} & \colhead{$r_{8102}$} & \colhead{$r_{8695}$} & \colhead{$r_{10800}$}}
\startdata
AA Tau\dotfill  & 061130 & 0.1  & 0.1  & 0.1  & 0.1  & 0.1  & 0.1  & 0.1  & 0.0  & 0.1  & 0.1  \\
AS 353A\dotfill & 061130 & 2.7  & 3.3  & 2.9  & 2.4  & 2.0  & 2.0  & 2.2  & 2.3  & 2.0  & 1.7  \\
AS 353A\dotfill & 061201 & \nod & \nod & \nod & \nod & \nod & \nod & \nod & \nod & \nod & 1.4  \\
BM And\dotfill  & 061130 & 0.3  & 0.3  & 0.2  & 0.2  & 0.3  & 0.2  & 0.2  & 0.2  & \nod & 0.5  \\
BP Tau\dotfill  & 061130 & 1.0  & 1.0  & \nod & 0.8  & 0.6  & 0.6  & 0.6  & 0.4  & 0.4  & 0.2  \\
CW Tau\dotfill  & 061130 & 0.6  & 0.9  & 0.8  & 0.7  & 0.7  & 0.8  & 0.8  & 0.7  & 0.8  & 1.7  \\
CW Tau\dotfill  & 061201 & 1.0  & 1.2  & 1.1  & 0.8  & 0.9  & 0.9  & 0.9  & 1.0  & 1.2  & 0.7  \\
CY Tau\dotfill  & 061130 & 0.7  & 1.0  & 0.8  & 0.5  & 0.7  & 0.6  & 0.2  & 0.0  & 0.2  & 0.2  \\
DF Tau\dotfill  & 061130 & 2.0  & 3.0  & 1.7  & 1.1  & 1.7  & 1.5  & 0.9  & 0.4  & 0.6  & 0.0  \\
DG Tau A\dotfill  & 061130 & 1.7  & 2.0  & 1.4  & 1.0  & 1.0  & 0.9  & 0.8  & 0.9  & 1.1  & 0.5  \\
DK Tau A\dotfill  & 061130 & 1.0  & 1.0  & 0.7  & 0.7  & 0.7  & 0.6  & 0.5  & 0.6  & 0.4  & 0.4  \\
DL Tau\dotfill  & 061130 & 2.0  & 2.0  & 1.3  & 1.0  & 1.2  & 1.3  & 1.1  & 1.4  & 1.0  & 1.0  \\
DO Tau\dotfill  & 061201 & 2.5  & 2.3  & 2.0  & 2.0  & 1.2  & 1.1  & 0.8  & 0.8  & 0.6  & 0.3  \\
DR Tau\dotfill  & 061130 & 6.3  & 7.3  & 7.0  & 6.0  & 5.8  & 5.0  & 3.5  & 3.0  & 2.9  & 3.5  \\
HN Tau A\dotfill  & 061130 & 3.5  & 3.3  & 2.7  & 1.4  & 1.8  & 1.6  & 1.5  & 1.0  & 1.4  & (1.9)  \\
LkCa 8\dotfill  & 061130 & 0.6  & 0.6  & 0.5  & 0.4  & 0.4  & 0.2  & 0.2  & 0.1  & 0.1  & 0.1  \\
UY Aur A\dotfill  & 061201 & \nod & \nod & \nod & \nod & \nod & \nod & \nod & \nod & \nod & 0.3
\enddata
\tablecomments{Parentheses indicate an upper limit (see \S\ 3.2.1).}
\end{deluxetable*}

\begin{deluxetable*}{lccccccccccc}
\tablecaption{SpeX Line Veilings\label{t.spexveil}}
\tabletypesize{\footnotesize}
\tablewidth{0pt}
\tablehead{\colhead{Object} & \colhead{UT Date} & \colhead{$r_{0.82}$} & \colhead{$r_{0.91}$} & \colhead{$r_{0.97}$} & \colhead{$r_{1.05}$} & \colhead{$r_{1.18}$} & \colhead{$r_{1.31}$} & \colhead{$r_{1.98}$} & \colhead{$r_{2.11}$} & \colhead{$r_{2.20}$} & \colhead{$r_{2.26}$}}
\startdata
AA Tau\dotfill   & 061127 &  0.1 & 0.25 &  0.2 &  0.2 &  0.2 &  0.2 &  0.3 &  0.3 &  0.3 &  0.5 \\
AS 353A\dotfill  & 061126/27\tablenotemark{a} & \nod & \nod & \nod & \nod & \nod & \nod & \nod &  3.8 & \nod & \nod \\
BM And\dotfill   & 061127 &  0.3 &  0.0 &  0.2 &  0.4 &  0.5 &  0.5 & 1.4 &  1.7 &  1.8 &  2.0 \\
BP Tau\dotfill   & 061126 &  0.3 &  0.4 &  0.4 &  0.5 &  0.5 &  0.5 &  0.8 &  0.9 &  1.0 &  1.0 \\
CW Tau\dotfill   & 061126 &  1.3 & \nod &  1.3 &  0.5 &  0.9 &  1.6 &  4.5 &  4.4 &  8.1 &  5.7 \\
CY Tau\dotfill   & 061127 & 0.15 &  0.2 &  0.2 &  0.2 &  0.4 &  0.3 &  0.3 &  0.5 &  0.5 &  0.6 \\
DF Tau\dotfill   & 061126 &  0.5 &  0.3 &  0.4 &  0.5 &  0.8 &  0.7 &  0.7 &  1.1 &  1.1 &  1.1 \\
DG Tau A\dotfill   & 061126 &  (2.3) & \nod &  1.3 &  0.9 &  0.8 &  1.5 &  1.7 &  1.2 &  1.2 &  2.1 \\
DK Tau A\dotfill   & 061126 &  0.6 &  0.6 &  0.5 &  0.5 &  0.7 &  0.6 &  1.5 &  1.9 &  2.0 &  2.0 \\
DL Tau\dotfill   & 061127 &  1.8 & \nod &  1.7 &  1.8 &  2.9 &  1.7 &  2.5 &  2.3 &  3.0 &  3.1 \\
DO Tau\dotfill   & 061127 &  0.7 & \nod &  0.6 &  0.7 &  1.1 &  1.3 &  2.2 &  2.6 &  3.1 &  2.9 \\
DR Tau\dotfill   & 061127 &  3.0 & \nod &  2.5 &  3.5 & \nod &  6.0 & \nod &  9.0 &   10 &   10 \\
HN Tau A\dotfill   & 061127 &  1.1 &  1.0 &  0.8 &  1.0 &  1.4 &  1.5 &  3.2 &  2.4 &  5.1 &  3.2 \\
LkCa 8\dotfill   & 061127 &  0.0 & \nod &  0.1 &  0.1 &  0.2 &  0.2 &  0.5 &  0.6 &  0.6 &  0.7 \\
RW Aur A\dotfill & 061126 & \nod & \nod & \nod & \nod & \nod & \nod & \nod &  3.9 & \nod &  6.7 \\
UY Aur A\dotfill   & 061127 &  0.5 & \nod &  0.3 &  0.5 &  0.8 &  0.8 &  1.4 &  1.4 &  1.9 &  1.7
\enddata
\tablecomments{Parentheses indicate an upper limit (see \S\ 3.2.1).}
\tablenotetext{a}{Two nearly identical spectra of 2006 Nov 26 and 27 were averaged to increase the signal-to-noise ratio.}
\end{deluxetable*}

Spectral templates for the line-veiling measurements include the weak T Tauri star (WTTS) V819 Tau (spectral type K7; \citealt{ken95}) and a grid of dwarf standards obtained with the same instrumental setup. The measurements of $r_\lambda$ in the echelle spectra (HIRES and NIRSPEC) are shown in Figure~\ref{f.echelle}, along with the polynomial fits $V_\lambda$ to characterize the behavior with wavelength. In the optical spectra, veiling is measured over a full order for those orders with strong photospheric lines, no line emission, and no problems with corrections for telluric absorption, which are particularly acute at longer wavelengths. Although there are 38 optical orders averaging about 100 \AA\ per order, the number of orders measured per star is considerably less than this and is not uniform among all the stars.  Also, the order-to-order scatter is small in many cases, but in a few it is considerable.  For the simultaneous NIRSPEC spectra, each of the 14 orders is about 120 \AA\ wide, and again, not all yield reliable veiling measurements. In the $Y$ band, we find no appreciable changes in veiling with wavelength, so only a single value is plotted at 1.08~\micron\ in Figure~\ref{f.echelle}.  Echelle veilings from both spectrographs are tabulated for nine wavelengths in Table~\ref{t.hiresveil}, where each HIRES value is an average of three adjacent orders (about 300 \AA) and the single NIRSPEC value is an average of several adjacent orders but is representative of the veiling over a range of about 1700 \AA.

In the wavelength coverage of the echelle data, from 0.48 to 1.1 \micron, most stars show a monotonic decrease in veiling from the bluest wavelength down to 0.65 \micron.  Beyond that, some continue to decrease to longer wavelengths, some flatten out to a constant veiling approaching the 1~\micron\ region, and two (DG Tau A and a second observation of CW Tau, shown in gray) rise toward 1~\micron. The two stars with the lowest veilings, AA Tau and BM And, show no convincing wavelength dependence to the veiling through the optical region, and none of the stars show convincing differences in veiling through the $Y$ band. The resulting spectra for the excess emission $E_\lambda$ will be shown in Section 3.3 along with the results from the SpeX data.

\subsection{SpeX Spectra}

\subsubsection{SpeX Line Veilings}

For the lower-resolution SpeX spectra, line veilings $r_\lambda$ provide the basis for disentangling the effects of extinction and veiling on the relatively flux-calibrated spectra, directly yielding the spectrum of the excess continuum emission $E_\lambda$ and the continuous veiling function $V_\lambda$.   

\begin{figure*}
\plotone{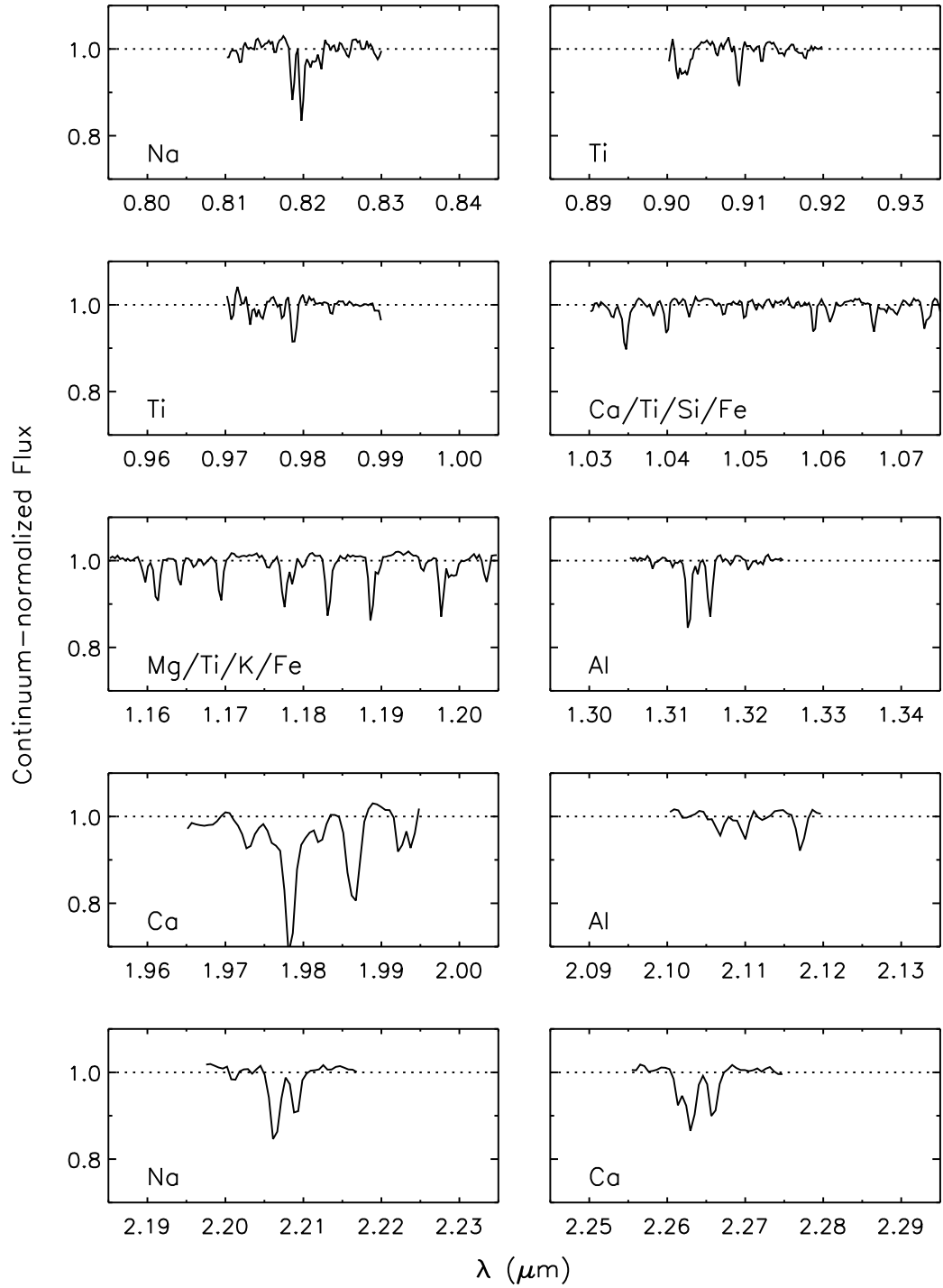}
\figcaption{The ten spectral regions with the strongest photospheric features used to measure veilings in the SpeX data, shown for the WTTS V819 Tau.  The dominant contributors to the absorption line spectrum in each region are indicated according to \citet{ray09}.\label{f.tenveil}}
\end{figure*}

The procedure to determine line veilings from the SpeX data is similar to that for the echelle spectra, although the process is hampered by the order-of-magnitude lower spectral resolution.   For spectral templates we have dwarfs and giants in the IRTF SpeX spectral library \citep{ray09} plus our own SpeX spectra of three WTTS:  HBC 407 (G8, observed on 2009 Dec 5), V819 Tau (K7, observed on 2006 Nov 26), and LkCa 14 (M0, observed on 2009 Dec 30). We identified ten spectral segments from 200 to 500 \AA\ wide that cover the strongest photospheric features, illustrated in Figure~\ref{f.tenveil} for V819 Tau. At SpeX resolution many of the features are unresolved blends, but the strongest contributors to each region are identified in the figure. We found the WTTS provided better templates for line veilings than dwarf templates from the IRTF library as judged from a smaller scatter in $r_\lambda$ with wavelength.  We thus used HBC 407 for BM And and V819 Tau for the rest of the sample.

The veilings measured for each of the ten wavelength regions are listed in Table~\ref{t.spexveil}.   For some of the high-veiling CTTS, either line or continuum emission made it impossible to determine veilings in all wavelength segments. The most severe cases were AS 353A and RW Aur A, where photospheric features could be discerned in only a few of the longer-wavelength regions near 2~\micron.  In the spectrum of DG Tau A, the wavelength regions shortward of 1 \micron\ contain some photospheric features that are in emission.  Although the specific lines used for veiling measurements display absorption, the adjacent line emission likely compromises the measurements in these regions.

The presentation of the combined veilings from the echelle and SpeX data is in \S\ 3.5, after $V_\lambda$ has been derived from the excess emission. Since this process depends on the low-resolution SpeX veilings, we need to assess their accuracy. We do this by comparing them to the echelle veilings in the region of spectral overlap.  Differences could arise either from error in the SpeX veilings or from veiling variations in the 3--4 day interval between the two sets of observations. As a check on variability, we can also compare emission-line equivalent widths in the region of spectral overlap. 

\begin{figure*}
\plotone{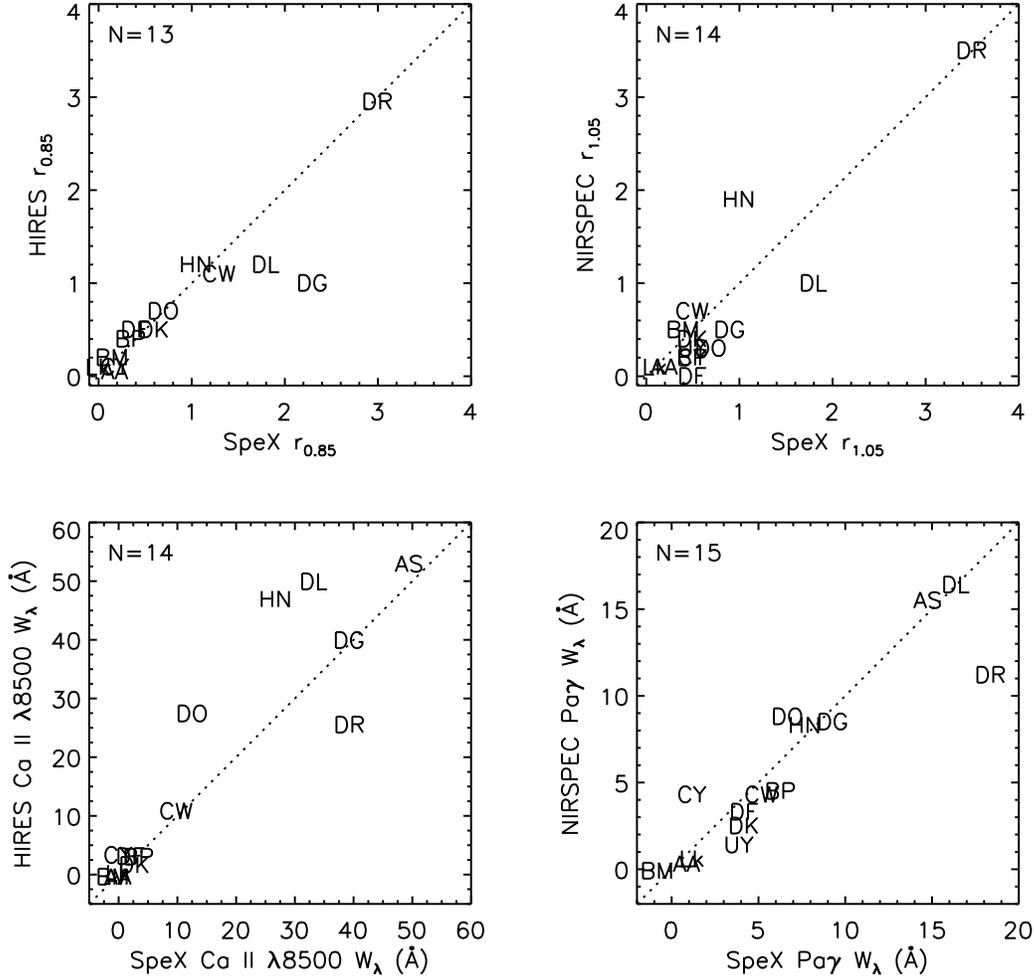}
\figcaption{Comparison of veilings and emission-line equivalent widths from SpeX and HIRES+NIRSPEC, separated by 3--4 days, in the region of overlap.  {\em Top left:} Veiling near 0.85 \micron\ in the HIRES versus the SpeX data. {\em Top right:} Veiling near 1.05 \micron\ in the NIRSPEC versus the SpeX data. {\em Bottom left:} Equivalent width of the \ion{Ca}{2} $\lambda8500$ line in the HIRES versus the SpeX data. {\em Bottom right:} Equivalent width of the Pa$\gamma$ line in the NIRSPEC versus the SpeX data.\label{f.linevar}}
\end{figure*}

These comparisons are made in Figure~\ref{f.linevar}.  The veiling comparisons are shown in the two upper panels, between SpeX and HIRES at 0.85 \micron\ on the left and between SpeX and NIRSPEC at 1.05 \micron\ on the right. The  line equivalent widths are compared in the two lower panels, between SpeX and HIRES for \ion{Ca}{2}~$\lambda$8500 on the left and between SpeX and NIRSPEC for Pa$\gamma$ on the right. These equivalent widths, along with those for Br$\gamma$ from the SpeX data, are listed in Table~\ref{t.linex}. (The number of stars in each panel is less than the full 16 because two stars were not observed with HIRES, one was not observed with NIRSPEC, and shorter-wavelength SpeX veilings could not be measured in two stars.)  

\begin{deluxetable*}{lccccc}
\tablecaption{Emission-Line Equivalent Widths\label{t.linex}}
\tabletypesize{\footnotesize}
\tablewidth{0pt}
\tablehead{\colhead{} & \colhead{SpeX} & \colhead{HIRES} & \colhead{SpeX} & \colhead{NIRSPEC} & \colhead{SpeX} \\ \colhead{} & \colhead{\ion{Ca}{2} $\lambda$8500} & \colhead{\ion{Ca}{2} $\lambda$8500} & \colhead{Pa$\gamma$} & \colhead{Pa$\gamma$} & \colhead{Br$\gamma$} \\ \colhead{Object} & \colhead{(\AA)} & \colhead{(\AA)} & \colhead{(\AA)} & \colhead{(\AA)} & \colhead{(\AA)}}
\startdata
AA Tau\dotfill   &   0.0 & $-$0.4 &  0.9 &  0.3 &  1.6 \\
AS 353A\tablenotemark{a}\dotfill  &  49.6 & 52.9 & 14.8 & 15.5 & 17.9 \\
AS 353A\dotfill  &  53.2 &  \nod & 18.4 & 14.7 & 21.3 \\ 
BM And\dotfill   &  $-$0.9 & $-$0.4 & $-$0.8 & $-$0.1 &  0.9 \\
BP Tau\dotfill   &   3.5 &  3.0 &  6.3 &  4.5 &  4.7 \\
CW Tau\tablenotemark{a}\dotfill   &   9.9 & 10.8 &  5.2 &  4.3 &  3.3 \\
CW Tau\dotfill   &   \nod & 11.0 &  \nod &  5.7 &  \nod \\
CY Tau\dotfill   &   0.0 &  3.3 &  1.2 &  4.3 &  1.1 \\
DF Tau\dotfill   &   2.1 &  3.1 &  4.2 &  3.3 &  3.3 \\
DG Tau A\dotfill   &  39.3 & 39.9 &  9.3 &  8.5 &  7.4 \\
DK Tau A\dotfill   &   2.8 &  1.6 &  4.2 &  2.5 &  2.2 \\
DL Tau\dotfill   &  33.3 & 49.9 & 16.4 & 16.4 & 12.0 \\
DO Tau\dotfill   &  12.6 & 27.4 &  6.7 &  8.8 &  2.5 \\
DR Tau\dotfill   &  39.4 & 25.5 & 18.4 & 11.2 &  8.6 \\
HN Tau A\dotfill   &  26.8 & 46.9 &  7.7 &  8.3 &  4.3 \\
LkCa 8\dotfill   &   0.0 &  0.2 &  1.2 &  0.6 &  1.5 \\ 
RW Aur A\dotfill &  69.6 &  \nod & 13.7 &  \nod & 10.2 \\
UY Aur A\dotfill   &   2.0 &  \nod &  3.9 &  1.4 &  2.0
\enddata
\tablenotetext{a}{AS 353A and CW Tau were observed twice by at least one instrument. The data from the epochs marked with this symbol are used in all plots.}
\end{deluxetable*}

In general we find good agreement between the veilings measured with the echelle and SpeX spectra taken a few days apart, although the correspondence is better between HIRES and SpeX than between NIRSPEC and SpeX for the majority of stars. Only three stars, all with high veilings, show significant veiling differences, up to a factor of 2, at these two wavelengths.  One of them, DL Tau, shows higher veiling with SpeX than in the echelle spectra at both 0.85 and 1.05 \micron, while the other two show discrepancies at only one wavelength. (DG Tau A has higher veiling at 0.85 \micron\ with SpeX than with HIRES, and HN Tau A has higher veiling at 1.05 \micron\ with NIRSPEC than with SpeX.) As will be apparent when we introduce the full spectrum of the veiling $V_\lambda$ in \S\ 3.5, the excess emission of DL Tau clearly varied over a three-day interval, but for the other two stars the discrepant points are anomalously high compared to echelle and SpeX veilings at other wavelengths.  We thus attribute them to error, possibly from line emission filling in photospheric features, and consider them to be upper limits. The line equivalent widths at the two epochs, which are not as sensitive to the difference in spectral resolution, are again similar for most stars, although \ion{Ca}{2}~$\lambda$8500 is more variable on a three-day timescale than Pa$\gamma$. We conclude that the lower-resolution SpeX line veilings in Table~\ref{t.spexveil} are generally reliable, and we use them to derive the spectra of the excess continuum emission in the next section. 

\subsubsection{Method for Finding the Spectrum of the Excess Continuum from SpeX Spectra\label{s.gullbring}}

For the relatively flux-calibrated SpeX data, we use line veilings to extract the broad SED of the excess continuum emission. We start with the assumption that the spectrum $O_\lambda$ of a CTTS observed with SpeX can be described as the reddened sum of the flux from the photosphere $P_\lambda$ and the flux from a continuum excess $E_\lambda$, such that  
\begin{equation}O_\lambda=\left(P_\lambda+E_\lambda\right)10^{-0.4A_{\lambda,O}}= P_\lambda\left(1+r_\lambda\right)10^{-0.4A_{\lambda,O}} \label{e.a}.\end{equation}

To find $E_\lambda$ we apply the method of \citet{gul98a}, used to derive excess spectra from medium-resolution optical spectrophotometry of CTTS.  The intrinsic CTTS photosphere $P_\lambda$ will differ from that of a spectral template $T_\lambda$ of identical temperature with extinction $A_{\lambda,T}$ by a wavelength-independent scaling factor $C$ that reflects the imprecision in the absolute flux calibration (\S~\ref{s.spex}) as well as the different distances and radii of the photospheric standards and the program stars: \begin{equation}P_\lambda=C T_\lambda 10^{0.4A_{\lambda,T}}.\end{equation}

The extinction to the program star is then determined by evaluating equation~(\ref{e.a}) at the set of wavelengths where individual line veilings $r_\lambda$ have been measured. Via substitution, equation~(\ref{e.a}) can be rewritten as 
\begin{eqnarray}O_\lambda&=&C T_\lambda\left(1+r_\lambda\right)10^{-0.4\left(A_{\lambda,O}-A_{\lambda,T}\right)}\nonumber\\
&=&C T_\lambda\left(1+r_\lambda\right)10^{-0.4k_\lambda\left(A_{V,O}-A_{V,T}\right)},
\end{eqnarray}
where the extinction law $k_\lambda=A_\lambda/A_V$ is assumed to be the same for both objects.  Here we use the extinction law of \citet{fit99} with $R_V=3.1$, as represented in the routine {\em fm\_unred.pro} in the IDL Astronomy Library.\footnote{http://idlastro.gsfc.nasa.gov/}

Reformatting by rearranging terms, taking the logarithm of both sides, and multiplying by 2.5 yields
\begin{equation}2.5\log\left[T_\lambda\left(1+r_\lambda\right)/O_\lambda\right]=k_\lambda\left(A_{V,O}-A_{V,T}\right)-2.5\log C,\label{e.d}\end{equation} which is identical in form to equation~(5) of \citet{gul98a}.  This equation is linear in $k_\lambda$, such that a line described by it has slope $A_{V,O}-A_{V,T}$ and intercept $-2.5\log C$. Thus both of these are readily determined from a linear fit to $2.5\log\left[T_\lambda\left(1+r_\lambda\right)/O_\lambda\right]$ versus $A_\lambda/A_V$. Once $A_{V,O}$, the extinction to the CTTS, is found, is it straightforward to recover the spectrum of the excess emission from the reddened and veiled $O_\lambda$.

\begin{figure*}
\plotone{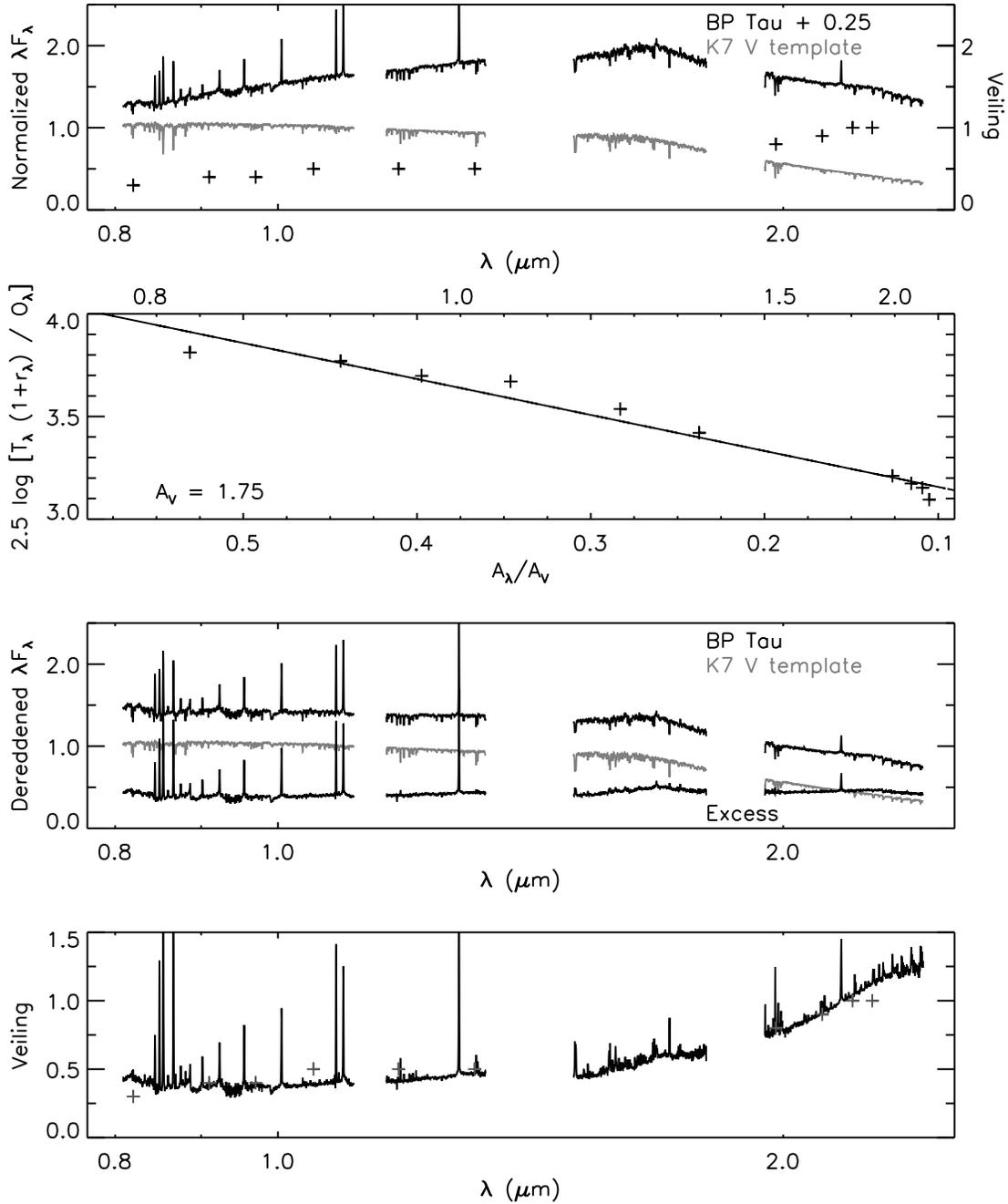}
\figcaption{Method for deriving $A_V$, the excess emission spectrum $E_\lambda$, and the continuum veiling $V_\lambda$ with SpeX data.  {\em Top panel:} Observed spectra of the {\it object} BP Tau (black and offset for clarity), the {\it template} K7 dwarf HD 237903 (gray), and the SpeX line veilings $r_\lambda$ for BP Tau ($+$).  {\em Second panel:} Veiling-corrected logarithm of the template-to-object flux ratio at each wavelength with a measured $r_\lambda$, plotted against the ratio of $A_\lambda$ to $A_V$ for the adopted reddening law (see eqn.\ 4). The slope of the best-fit line is $A_{V,O}-A_{V,T}$. {\em Third panel}: Normalized, reddening-corrected spectra of BP Tau (black), the K7 V template (gray), and their difference (black), i.e., the excess spectrum $\lambda E_\lambda$.  {\em Bottom panel}: The continuous veiling $V_\lambda$ (solid curve), resulting from dividing $E_\lambda$ by the template, and the line veilings $r_\lambda$ ($+$).\label{f.bptau}}
\end{figure*}

The key steps in the above process of extracting the SED of the excess emission are illustrated in Figure~\ref{f.bptau} for BP Tau. The upper panel shows the observed spectra of both BP Tau, $\lambda O_\lambda$, and the K7 V photospheric template HD 237903 from the IRTF library, $\lambda T_\lambda$, plus the ten measured line veilings $r_\lambda$.  The spectra are plotted in units of $\lambda F_\lambda$ scaled to unity at 0.8~\micron, although the spectrum of BP Tau is shifted upward by 0.25 for clarity.  The second panel plots $2.5\log\left[T_\lambda\left(1+r_\lambda\right)/O_\lambda\right]$ against $k_\lambda$ at each of the ten wavelengths for which the veiling has been measured.  As indicated in equation~(\ref{e.d}), this should be a linear relation, with a slope equal to $A_{V,O}-A_{V,T}$.  Since the template HD~237903 has $A_{V,T}=0$ \citep{ray09}, the slope of the best-fit line is the extinction toward the object $A_{V,O}$.  (Note that, because $k_\lambda$ increases from right to left, a line that increases from right to left represents a positive $A_{V,O}$.)  Once the extinction is found, the observed spectrum $\lambda O_\lambda$ can be dereddened, as shown in the third panel.  If the template spectrum is then multiplied by $C$ (the normalization constant corresponding to the intercept of the best-fit line), it can be subtracted from the dereddened object, yielding the excess spectrum $\lambda E_\lambda$, also shown in panel 3.  The template spectrum in panel 3 is scaled such that $\lambda F_\lambda=1$ at 0.8~\micron, and the other two spectra are shown with the correct scaling relative to the template. We will express the excess $\lambda E_\lambda$ in units of the photospheric continuum at 0.8 \micron\ in all subsequent figures and tables.

Finally, the continuous veiling spectrum $V_\lambda$, shown in panel 4, is found by dividing the spectrum of the excess emission $E_\lambda$ by the scaled photospheric template $P_\lambda$.  We compare it to the measured line veilings $r_\lambda$ as a consistency check. If the line veilings lie close to the derived veiling spectrum, as seen for BP Tau in Figure~\ref{f.bptau}, then the template is a good match to the CTTS photosphere, and $A_V$ is well determined. 

The normalization constant $C$ is related to the radius of the T Tauri star if the absolute flux calibration is accurate. Provided the spectra of the template and the dereddened, deveiled CTTS have the same wavelength dependence, their flux ratio can be written $C=\epsilon(R_{\rm CTTS}/d_{\rm CTTS})^2/(R_{\rm temp}/d_{\rm temp})^2$, where $\epsilon$ accounts for errors in the absolute flux calibration.  Solving for the CTTS radius, $R_{\rm CTTS}=(C/\epsilon)^{1/2}\left(d_{\rm CTTS}/d_{\rm temp}\right)R_{\rm temp}$.  With BP Tau as an example and HD 237903 as its template, $C=6.42\times10^{-2}$; $d_{\rm temp}=12.9$~pc \citep{gou04}; $d_{\rm CTTS}=140$~pc, a commonly accepted distance to Taurus \citep{ken08}; and $R_{\rm temp}=0.67~R_\sun$ \citep{joh83}, giving a radius of $1.85~R_\sun\pm20\%$ for BP Tau if the error is dominated by a 40\% uncertainty in the absolute flux calibration (\S~2.1).  This compares favorably with other estimates for the radius of this star: 1.9 $R_\sun$ \citep{har95}, 1.99 $R_\sun$ \citep{gul98a}, and 1.95 $R_\sun$ \citep{joh99}, indicating that either the SpeX absolute flux calibrations for BP Tau and its telluric standard are reasonably accurate, or that any inaccuracies cancel when the CTTS is divided by its telluric standard.

Before showing the results of this procedure, summarized in Figure~\ref{f.bptau}, to find $A_V$, the spectrum of the excess emission $E_\lambda$, and the spectrum of the veiling $V_\lambda$ for each star in the SpeX data set, we demonstrate how the outcome depends on the properties of the chosen extinction law and spectral templates.

\subsubsection{Sources of Uncertainty in Determination of  $A_V$ and $E_\lambda$ from SpeX Spectra}

\begin{figure*}
\epsscale{0.99}
\plotone{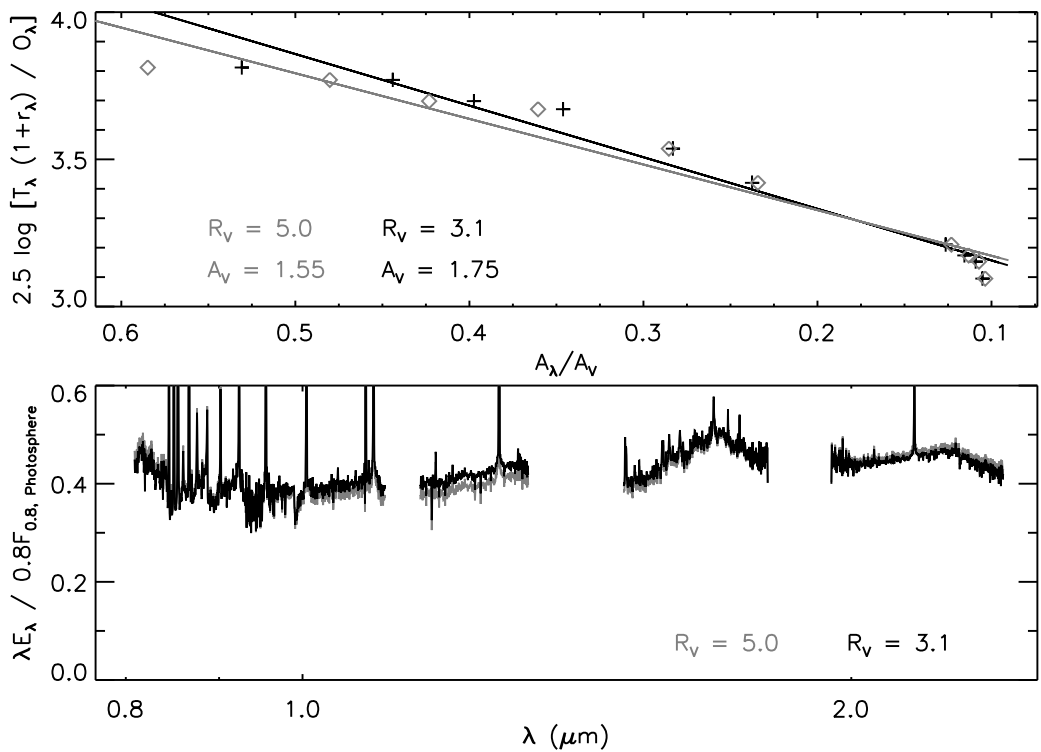}
\figcaption{Comparison of extinction laws with different $R_V$ in the derivation of the excess spectrum $E_\lambda$ of BP Tau. {\em Top panel:}  Veiling-corrected logarithm of the template-to-object flux ratio at each wavelength with a measured $r_\lambda$, plotted against the ratio of $A_\lambda$ to $A_V$ for each extinction law. The slope of the linear relation is  $A_{V,O}-A_{V,T}$ (eqn.\ 4). {\em Bottom panel}:  The corresponding excess spectrum $\lambda E_\lambda$ for each extinction law. Although the derived $A_V$ is smaller for $R_V=5.0$, the difference in the derived $\lambda E_\lambda$ is not significant at SpeX SXD wavelengths.\label{f.bptauR5}}
\epsscale{1}
\end{figure*}

The extraction of the excess emission spectrum from our SpeX data depends on the assumed extinction law and the spectrum of the template.  We experimented with several extinction laws from \citet{fit99} and \citet{cal04} thought to be more appropriate for dark clouds than the standard ISM extinction with $R_V=3.1$. We compare in Figure~\ref{f.bptauR5} the effect on the derived $A_V$ and $E_\lambda$ between extinction laws with $R_V=5.0$ and $R_V=3.1$ \citep{fit99}.  The larger $R_V$ yields a small decrease in $A_V$ of 0.2 mag (1.55 instead of 1.75, a 10\% effect) and an excess spectrum that differs from the $R_V=3.1$ result by an average of 3\% (up to 6\% at any one wavelength).  The effects are small in the SpeX SXD regime, since the extinction laws  have nearly the same ratios of $A_\lambda / A_V$ from 1.2 to 2.5 \micron\ (0.27 at $J$, 0.17 at $H$, and 0.11 at $K$) and differ only modestly from 0.8 to 1.2 \micron.   We thus adopt the standard interstellar extinction law ($R_V=3.1$) with the consequence that $A_V$ may be overestimated by a few tenths of a magnitude; the derived spectrum of the excess emission is not affected significantly.

The derivation of $A_V$ and $E_\lambda$ are also affected by the choice of a spectral template.  Ideally the photospheric template $T_\lambda$ would be matched in both temperature and surface gravity to the CTTS $O_\lambda$ in equation 4, and its extinction would be perfectly known.  Thus a good grid of WTTS with well defined extinctions would be preferable templates, but we acquired only three SpeX spectra of WTTS, with spectral types G8, K7, and M0. An alternative is to use the IRTF library, where we can select unreddened dwarf templates to match the spectral type of each program star. 

\begin{figure*}
\epsscale{0.99}
\plotone{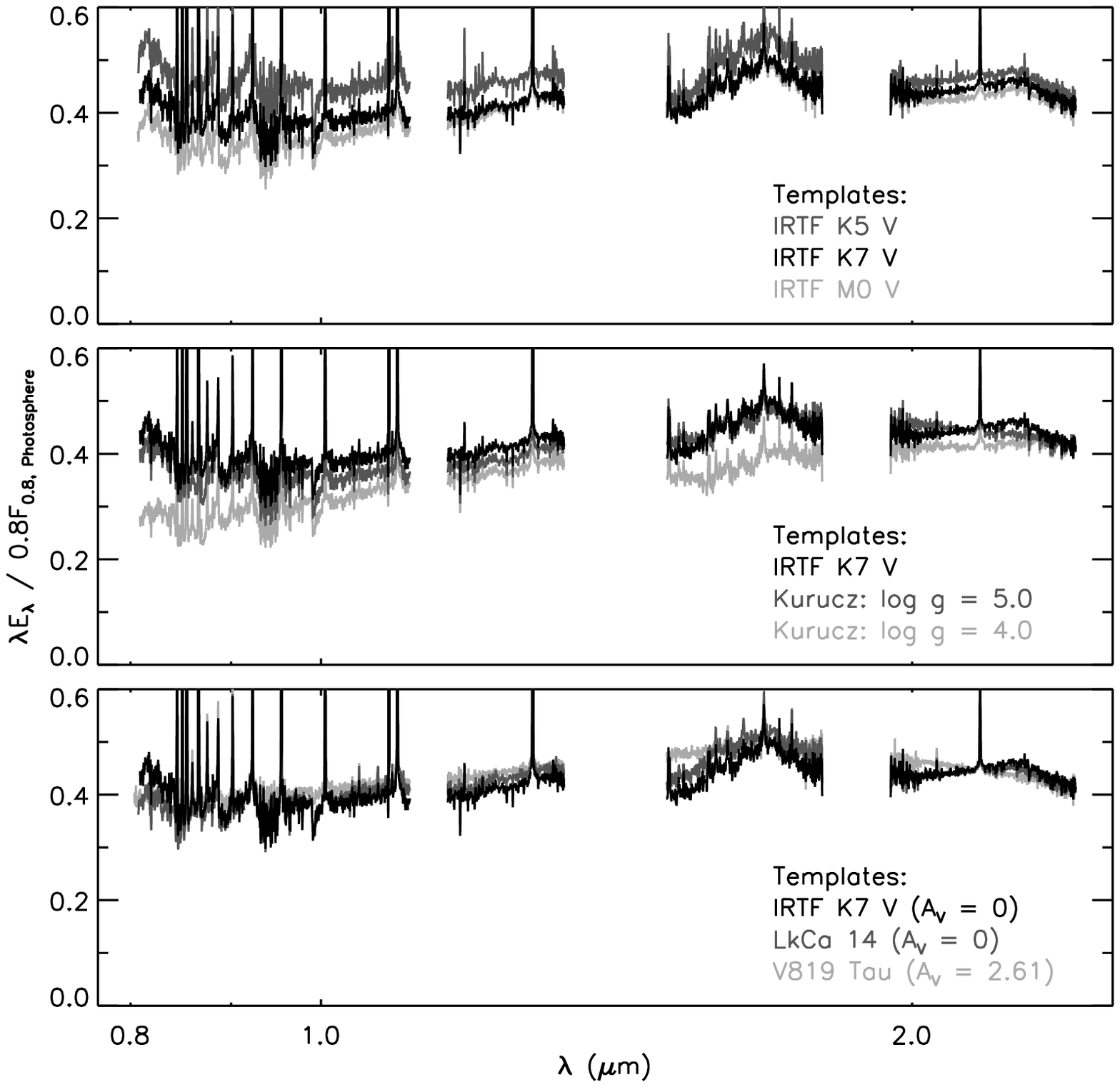}
\figcaption{The excess spectrum $\lambda E_\lambda$ of BP Tau (K7) for different choices of template.  {\em Top:} $\lambda E_\lambda$ based on IRTF K5 V, K7 V, and M0~V standards.  The resulting $A_V$ for BP Tau are 2.17, 1.75, and 1.48, respectively.  {\em Middle:} $\lambda E_\lambda$ repeated for the same K7 V standard and for two Kurucz models (adjusted to represent SpeX resolution; see text) with $T=4000$~K and $\log g=5.0$ and 4.0.  The resulting $A_V$ for BP Tau with these Kurucz models as templates are 1.58 and 1.05, respectively.  {\em Bottom:} $\lambda  E_\lambda$ repeated for the same K7 V standard and for two WTTS, LkCa 14 (M0) and V819 Tau (K7), with resulting $A_V$ for BP Tau of 1.71 and 1.75, respectively.  The $\lambda E_\lambda$ are normalized such that the photospheric $\lambda F_\lambda=1$ at 0.8~\micron.\label{f.compare}}
\epsscale{1}
\end{figure*}

In Figure~\ref{f.compare} we illustrate the consequences in the derivation of $A_V$, $E_\lambda$, and $V_\lambda$ for BP Tau that result from different choices for the spectral template.  In the top panel, we show the effect of errors in the temperature classification of BP Tau (nominally K7) by plotting its $\lambda E_\lambda$ with three dwarf spectra from the IRTF library as templates: a K7 and the next hottest (K5) and next coolest (M0) dwarfs available in the library.  While spectral types of CTTS are known to $\pm1$ subclass (see references in Table~\ref{t.sample}) and those of the templates are known to $\pm0.65$ subclasses \citep{ray09}, the figure shows that a mismatch as large as two subclasses (K5 versus K7) yields an error in the excess spectrum of only about 10\% at the most discrepant wavelengths.  Although the $A_V$ derived vary substantially (2.17 when the K5 dwarf is used, 1.75 when the K7 is used, and 1.48 when the M0 is used) this uncertainty does not impact our results on the emission excess because we use a reddening-independent approach to find $E_\lambda$ from our optical spectra.

In the middle panel we show $\lambda E_\lambda$ from three templates with the same temperature and different surface gravities: (1) the K7~V from the IRTF library, (2) a Kurucz model with $\log g=5.0$ and $T=4000$~K, and (3) a Kurucz model with $\log g=4.0$ and $T=4000$~K. Since the Kurucz models\footnote{ftp://ftp.stsci.edu/cdbs/grid/ck04models/} have lower spectral resolution than the SpeX data, the continuum shape from each model was combined with the lines from the IRTF standard so photospheric features would subtract cleanly.  The comparison shows that similar excesses are derived from both the IRTF template and the $\log g=5.0$ model. The excess derived from the $\log g=4.0$ model is slightly smaller in the $J$ and $K$ bands and more markedly so in the $I$ and $H$ bands. In particular, the derived excess from the lower surface-gravity model is flat between 0.8 and 0.85 \micron\ and through the $H$ band, while for the higher surface-gravity templates the spectrum rises shortward of 0.85~\micron\ and shows a small hump at $H$. This comparison suggests that if the Kurucz $\log g=4.0$ models are more appropriate templates for CTTS, then our use of dwarf standards will introduce a small error into the derivation of $E_\lambda$.

In the bottom panel of Figure~\ref{f.compare}, we compare $\lambda E_\lambda$ for BP Tau using two WTTS templates with spectral types K7 and M0 to the result from the IRTF K7 V standard. One WTTS, LkCa 14 (M0), has zero extinction and yields a $\lambda E_\lambda$ for BP Tau that shows a much better match to the excess found with the K7 V template than with the Kurucz $\log g=4.0$ model. The other WTTS, V819 Tau, of spectral type K7, does not have negligible extinction. If we adopt $A_V=2.6$ for V819 Tau, which follows from dereddening it against either LkCa 14 or a K7~V (see next paragraph), the derived excess for BP Tau is again comparable to that from LkCa 14 and the IRTF K7 V dwarf, although it is flatter in the $I$, $H$, and $K$ bands, and it is not a good match to the excess from the Kurucz $\log g=4.0$ model. 

The comparison shown in Figure~\ref{f.compare} suggests that the IRTF library, offering a fine grid of dwarf spectral types with no extinction, provides a reasonable choice for our spectral templates.  A potential consequence of this approach is that the surface gravity of a dwarf, typically $\log g=4.5$, is somewhat higher than the $\log g=4.0$ found for four WTTS from high-resolution photospheric line ratios \citep{tak10}, but not out of line considering the results in Figure~\ref{f.compare} and the sizable scatter inferred for $\log g$ distributions of pre--main-sequence stars from colors and line ratios \citep{hil09}. Surprisingly, however, no matter which spectral template we use, we derive an extinction to BP Tau that is significantly larger than the $A_V$ of 0.5 found by \citet{gul98a}. Our values for $A_V$ range from 1.75 to 1.05 (see above and caption to Fig.~\ref{f.compare}),  all larger than the overestimate of a few tenths of a magnitude that would be expected if we adopted a dark-cloud extinction law.

The large $A_V$ we consistently find from comparing CTTS to spectral templates in the SpeX SXD spectral regime may be related to a peculiar behavior in WTTS first noted by \citet{gul98a} and \citet{gul98b}. They found that one of the WTTS templates we use here, V819 Tau, has anomalous colors, becoming increasingly redder with wavelength than a main-sequence star of comparable spectral type. The observed color anomalies in V819 Tau, also found in a number of other WTTS but not in our other SpeX template LkCa~14, lead to the curious result that $A_V$ derived from colors increases with wavelength, with $A_V=0.5$ from $B-V$ to $A_V=1.8$ from $V-J$. Indeed, the $A_V=2.6$ we found for V819 Tau when dereddening its SpeX spectrum against either the WTTS LkCa 14 or a K7 V template (both with no reddening) continues this trend.  Gullbring et al.\ considered possible causes of anomalous colors and favored the (untested) explanation that very large cool spots might produce the observed effect.  That V819 Tau has an unrecognized emission excess in the spectral region of our study is worth considering, since it was recently found with Spitzer to have a weak IR excess at wavelengths exceeding 10 \micron\ \citep{fur09}.
 
As will be shown the next section, the high $A_V$ derived from SpeX SXD data is not unique to BP Tau; it is also the case for most of our CTTS.  Fortunately, this does not have a significant effect on our derived excesses, since in the SpeX wavelength region the extinction correction is not large, and in the optical we use a reddening-independent method to recover the SED of the excess emission. We return to this topic in the discussion section.

We conclude from the good agreement between the $\lambda E_\lambda$ derived from the unreddened WTTS LkCa 14 and the unreddened K7 V shown in Figure~\ref{f.compare} that using unreddened dwarfs from the IRTF library matched in spectral type to each CTTS is a reasonable choice for spectral templates, a practice also adopted by \citet{esp10} in the near infrared.  However, the shape of the derived $\lambda E_\lambda$ in the $I$ and $H$ bands may be slightly affected if there is a gravity mismatch between the CTTS and the template. 

\subsubsection{Derivation of $A_V$ from SpeX Spectra}

\begin{figure*}
\includegraphics[angle=90,width=\hsize]{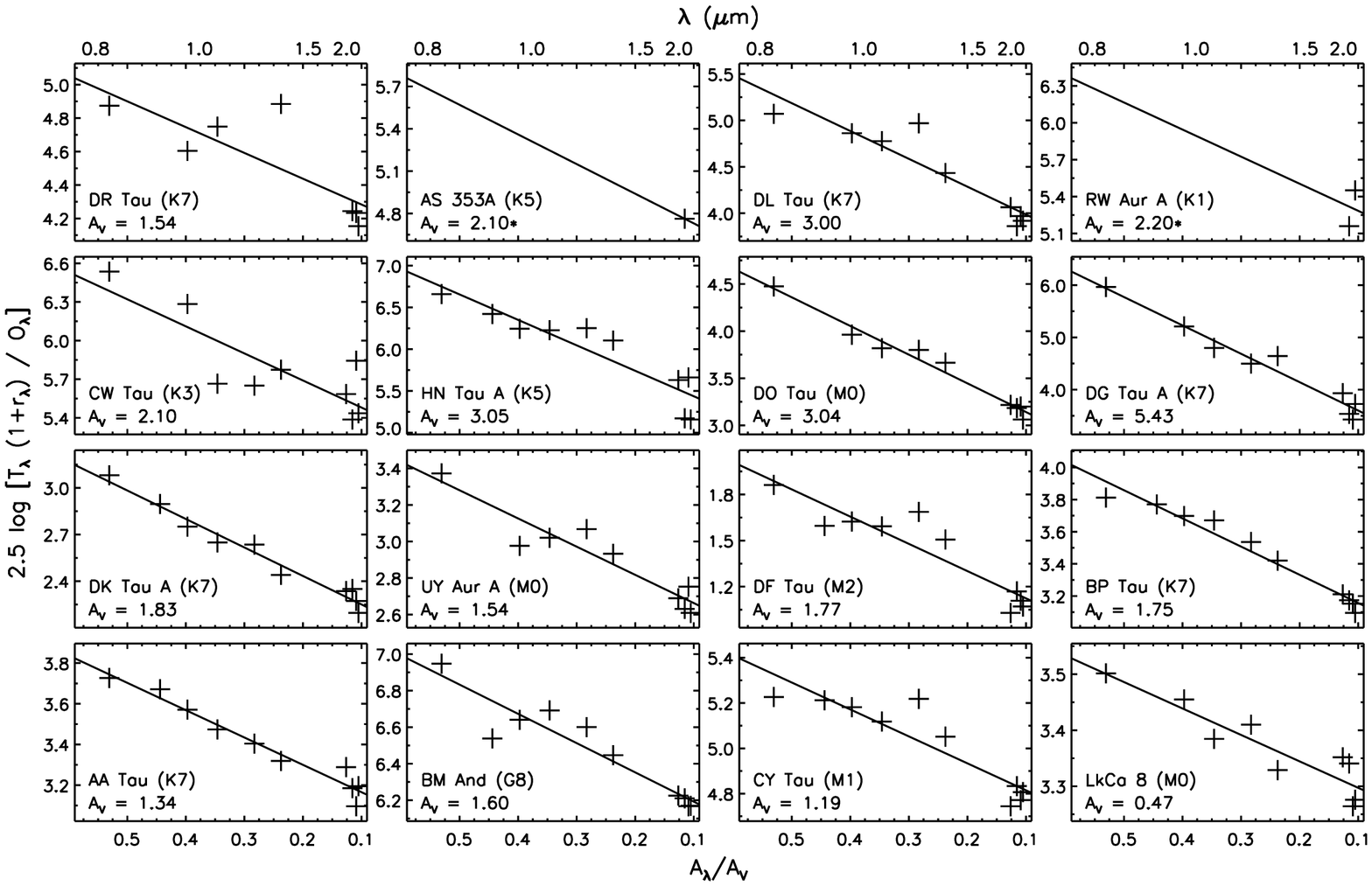}
\figcaption{Derivation of $A_V$ (same as the second panel of Fig.~\ref{f.bptau}) for each star in the sample.  The slope of the best-fit line to the veiling-corrected flux ratios (`+') is $A_V$, since in all cases the template has no extinction.  The two stars with an asterisk by their $A_V$ (AS~353A, RW Aur A) use published $A_V$ for the slopes, since there are not enough $r_\lambda$ points to derive them self-consistently.  The spectral type of the IRTF dwarf template used for each star is given in parentheses.  Stars are sorted by their 1 \micron\ excess. \label{f.gullbringa}}
\end{figure*}

In this section we derive the extinctions to the CTTS program stars from their SpeX spectra and observed line veilings following the method demonstrated in Figure~\ref{f.bptau}. For the reasons outlined in the previous section, we use IRTF dwarf standards with no extinction as spectral templates, and we use a standard extinction law.  The validity of these assumptions can be examined in Figure~\ref{f.gullbringa}, where we show the relation between $2.5\log\left[T_\lambda\left(1+r_\lambda\right)/O_\lambda\right]$ and $k_\lambda$ (see eqn.\ 4 and panel 2 in Fig.~\ref{f.bptau}), expected to be linear for each star.  For most stars there is little scatter around the best-fit line, indicating that the SpeX line veilings are robust. The slope of each line corresponds to the difference in extinction between the standard and the template, but since we have chosen IRTF standards with zero reddening, it gives $A_V$ for each star, as identified in each panel and listed in Table~\ref{t.fitresults}. The procedure had to be modified for the two stars for which only one or two veiling measurements could be determined due to line emission filling in the veiled photospheric features (AS 353A and RW Aur A). For these stars we used the published $A_V$ to set the slope of the line, anchoring it through the one or two valid veiling measurements.

\begin{deluxetable}{lccccc}
\tablecaption{Results of Fits to SpeX Line Veilings\label{t.fitresults}}
\tabletypesize{\footnotesize}
\setlength{\tabcolsep}{0.2in}
\tablewidth{0pt}
\tablehead{\colhead{Object} & \colhead{$A_{V\rm,obs}$} & {$A_{V\rm,lit}$} & \colhead{Ref} & \colhead{$C$} & \colhead{$R$ ($R_\sun$)} \\ \colhead{(1)} & \colhead{(2)} & \colhead{(3)} & \colhead{(4)} & \colhead{(5)} & \colhead{(6)}}
\startdata
AA Tau\dotfill   & 1.34 & 0.74 & 1 & $6.13\times10^{-2}$ & 1.81 \\
AS 353A\dotfill  & \nod\tablenotemark{a} & 2.1  & 2 & $1.56\times10^{-2}$ & 1.48 \\
BM And\dotfill   & 1.60 & 0.67 & 5 & $3.87\times10^{-3}$ & 2.55 \\
BP Tau\dotfill   & 1.75 & 0.51 & 1 & $6.42\times10^{-2}$ & 1.85 \\
CW Tau\dotfill   & 2.10 & 2.29 & 4 & $7.80\times10^{-3}$ & 1.53 \\
CY Tau\dotfill   & 1.19 & 0.32 & 1 & $1.32\times10^{-2}$ & 1.65 \\
DF Tau\dotfill   & 1.77 & 0.60 & 3 & $4.18\times10^{-1}$ & 3.40 \\
DG Tau A\dotfill & 5.43 & 3.2  & 2 & $6.00\times10^{-2}$ & 1.79 \\
DK Tau A\dotfill & 1.83 & 1.42 & 1 & $1.49\times10^{-1}$ & 2.82 \\
DL Tau\dotfill   & 3.00 & 1.7  & 2 & $3.35\times10^{-2}$ & 1.34 \\
DO Tau\dotfill   & 3.04 & 2.27 & 1 & $7.33\times10^{-2}$ & 1.77 \\
DR Tau\dotfill   & 1.54 & 3.2  & 2 & $2.23\times10^{-2}$ & 1.09 \\
HN Tau A\dotfill & 3.05 & 0.65 & 1 & $8.86\times10^{-3}$ & 0.78 \\
LkCa 8\dotfill   & 0.47 & 0.32 & 1 & $5.01\times10^{-2}$ & 1.46 \\
RW Aur A\dotfill & \nod\tablenotemark{a} & 2.2  & 2 & $9.43\times10^{-3}$ & 1.56 \\
UY Aur A\dotfill & 1.54 & 0.55 & 3 & $9.92\times10^{-2}$ & 2.06
\enddata
\tablecomments{Col.~2: Observed $A_V$; Col.~3: Literature $A_V$ from optical data; Col.~4: Reference for $A_V$; Col.~5: Scaling constant; Col.~6: Radius derived from scaling constant.}
\tablerefs{(1) \citealt{gul98a}; (2) \citealt{har95}; (3) \citealt{har03}; (4) \citealt{ken95}; (5) \citealt{ros99}.}
\tablenotetext{a}{Insufficient photospheric lines to determine $A_V$.}
\end{deluxetable}

One of our CTTS, AA Tau, is subject to periodic occultation by its circumstellar disk, and thus it may have anomalous extinction if seen through the disk.   Following the prescription of \citet{bou07}, in which the period of AA Tau is reported to be 8.22 days and JD 2453308 corresponds to phase 0.51, our Keck data were acquired at phase 0.21, and our IRTF data were acquired at phase 0.85. As photometric and spectroscopic diagnostics of disk occultation appear only between phases 0.3 and 0.8, it seems we are not sampling extinction through the disk. The low veiling we observe (0.1 in the optical and 0.3 at $K$) is consistent with the level found by Bouvier et al.\ outside of the occultation phase, further supporting this conclusion.

We also show in Table~\ref{t.fitresults} both the normalization constant $C$ that scales the template to the dereddened, deveiled CTTS and the corresponding stellar radius for each CTTS. As described in \S 3.3, $C$ depends on the radius of and distance to both the CTTS and the template. We adopted a distance to the Taurus objects of 140 pc, a distance to AS 353A of 200 pc \citep{ric06}, a distance to BM And of 440 pc \citep{ave69}, distances to templates other than HD 237903 \citep{gou04} from Hipparcos \citep{per97}, and template radii from \citet{joh83}.  The CTTS radii derived from $C$ lie between 0.8 and 3.4 $R_\sun$ with a mean of $1.8\pm 0.7~R_\sun$. As this is a typical CTTS radius, we infer that the absolute flux calibration of the SpeX data is reasonably accurate. 

The extinctions we derive from the SpeX data are typically 0.5 to 1 magnitude higher than those from published optical data, as shown in Table~\ref{t.fitresults}.  As described in the previous section, a modest overestimate of $A_V$ could come from using an extinction law that is not appropriate for dark clouds or from a spectral type mismatch, but these cannot account for the large differences found here.  Although our results are not compromised for the reasons given above, the tendency to derive larger $A_V$ in the SpeX regime for some WTTS and most CTTS is a subject for continued investigation, best addressed by a systematic campaign to acquire a dense grid of flux-calibrated WTTS spectra with wide wavelength coverage.

\subsection{The Veiling and Emission Spectra from\\0.48 \micron\ to 2.4 \micron}

In this section we present the spectra of the continuous veiling $V_\lambda$ and the continuous excess emission $E_\lambda$ obtained by merging the SpeX results with the HIRES and NIRSPEC results, thus covering the full wavelength range of our data, from 0.48 to 2.4 \micron.

\begin{figure*}
\plotone{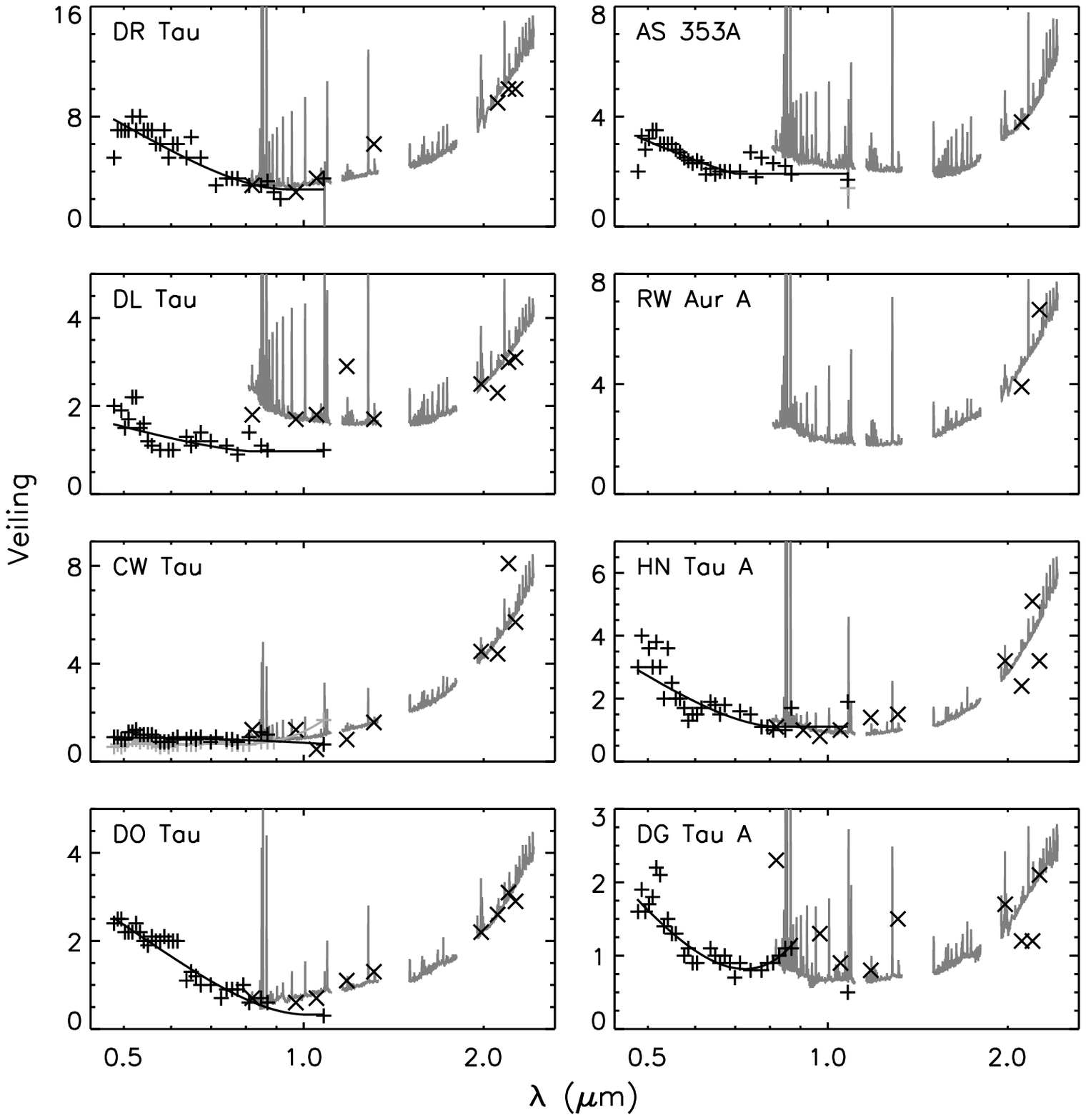}
\figcaption{Line-veiling measurements $r_\lambda$ compared to derived continuum veilings $V_\lambda$ over the wavelength range 0.48--2.4 \micron\ with stars sorted by their 1 \micron\ excess.  Points for $r_\lambda$ are ($+$) for HIRES and NIRSPEC and ($\times$) for SpeX.  Curves for $V_\lambda$ are black for HIRES and NIRSPEC (fits to individual $r_\lambda$) and dark gray for SpeX data (as in the bottom panel of Fig.~\ref{f.bptau}).  A second echelle spectrum for CW Tau is overplotted in light gray.\label{f.veilinga}}
\end{figure*}

The results for $V_\lambda$ are assembled in Figure~\ref{f.veilinga}, showing (1) the individual line veilings $r_\lambda$ from both the SpeX and echelle spectra, (2) the SpeX veiling curves $V_\lambda$ derived from the emission excesses, which include line emission, as in the fourth panel of Figure~\ref{f.bptau}, and (3) the echelle veiling curves $V_\lambda$ that are polynomial fits to the echelle veilings. In most cases there is reasonable agreement in the region of spectral overlap between the low- and high-resolution data, taken several days apart. The exception is DL Tau, where the SpeX $V_\lambda$ is about twice as high as the veiling from the echelle data; we attribute this difference to variability (see \S\ 3.2). 

\addtocounter{figure}{-1}
\begin{figure*}
\epsscale{1}
\plotone{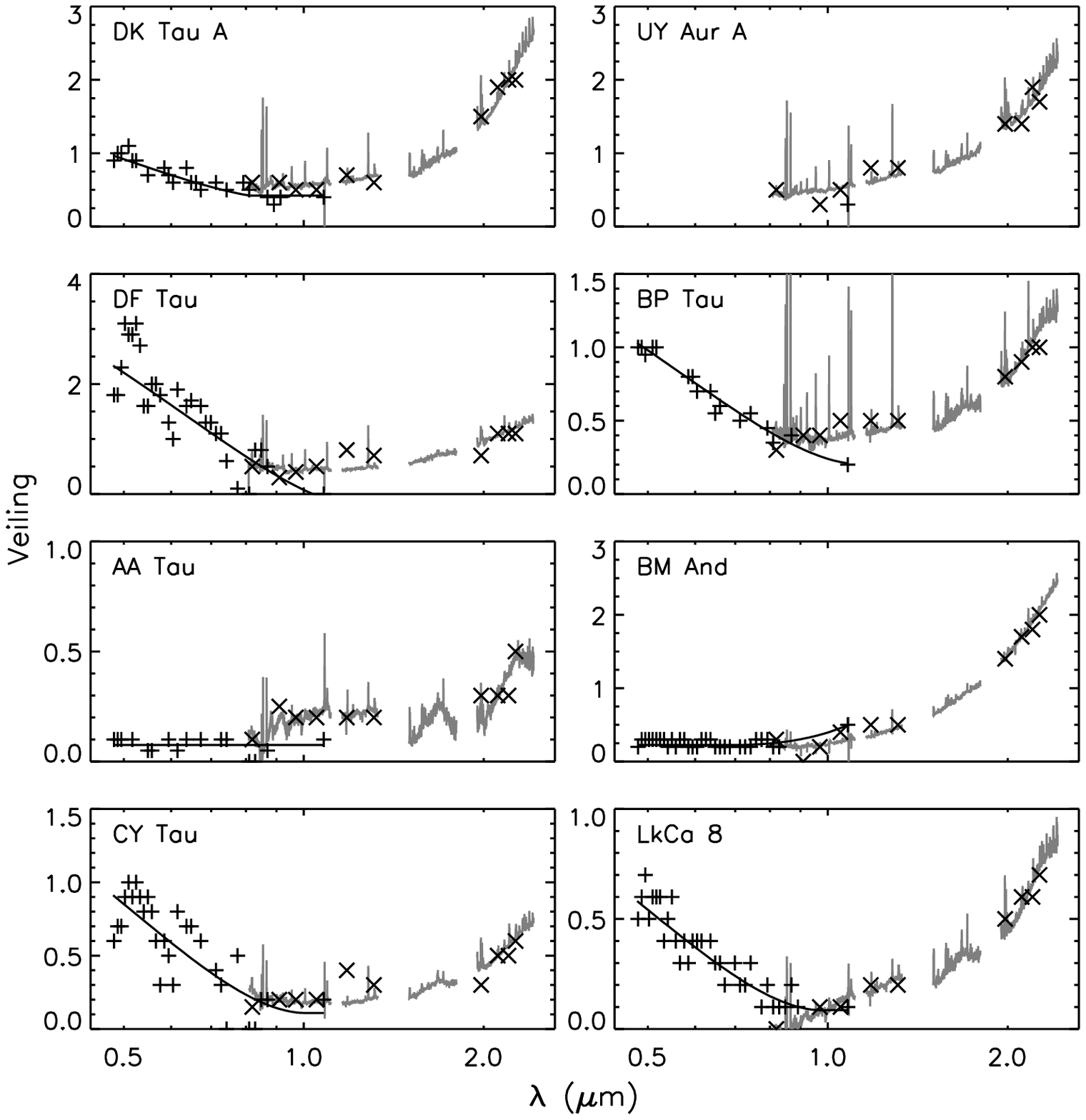}
\figcaption{cont.\label{f.veilingb}}
\end{figure*}

\begin{figure*}
\plotone{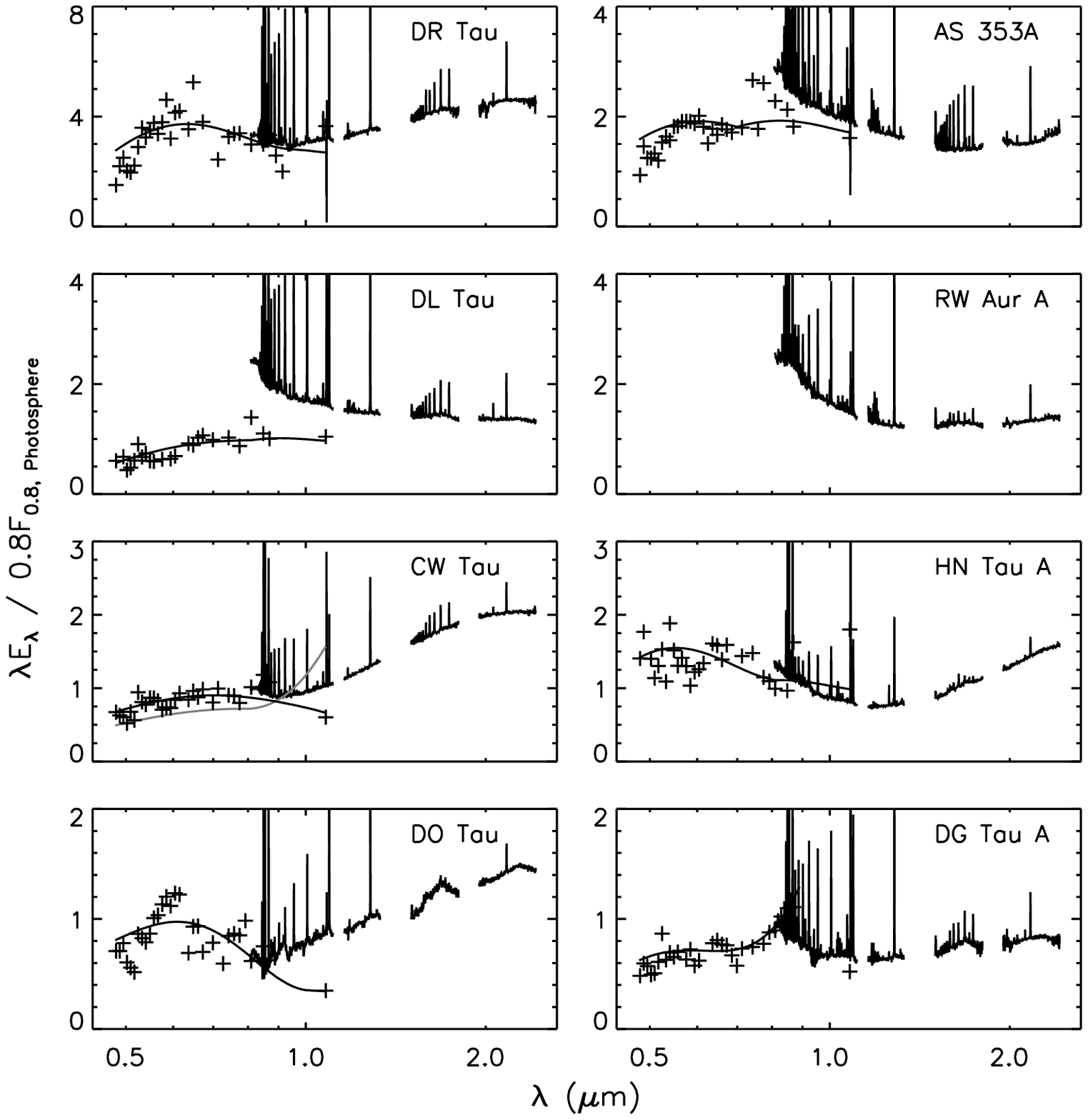}
\figcaption{Emission excess curves $\lambda E_\lambda$ over the wavelength range 0.48--2.4 \micron\ with stars sorted by their 1 \micron\ excess.  The SpeX excesses, which include emission lines, are calculated as in the third panel of Fig.~\ref{f.bptau}. For the HIRES and NIRSPEC excesses, the solid curve is the product of the fit to the line veilings $r_\lambda$ and the fit to the continuum of the photospheric template, and plus signs are point-by-point products of each individual line veiling and the flux in the template at that same wavelength.  Excesses are in units of the photospheric $\lambda F_\lambda$ at 0.8~\micron.  A second echelle spectrum for CW Tau is overplotted in gray.\label{f.excessa}}
\end{figure*}

The resulting SEDs of the continuum excess emission are illustrated in Figure~\ref{f.excessa}, plotted as $\lambda E_\lambda$ in units of the photospheric flux at 0.8 \micron. Over the range observed with SpeX, between 0.80 and 2.43 \micron, the $E_\lambda$ curves are generated with the procedure discussed in \S~\ref{s.gullbring} and illustrated in the third panel of Figure~\ref{f.bptau}, which yields emission lines as well as the continuum shape.  For the echelle data, between 0.48 and 1.08 \micron, the $\lambda E_\lambda$ curves are the product of polynomial fits to the line veilings (Fig.~\ref{f.veilinga}) and polynomial fits to the continua of temperature-matched dwarf standards from the Pickles library \citep{pic98}; this method is independent of extinction. For comparison, we also show (plus signs) the point-by-point products of individual veiling measurements and the corresponding Pickles template at the same wavelength. This shows the scatter in the individual line-veiling measurements, including fluctuations that would result from measuring within broad molecular bands. In most cases the individual measurements show little scatter around the curve. However, for the few stars that show substantial scatter (e.g., CY Tau and DF Tau), the curves for $\lambda E_\lambda$ may be less accurate.  (The scatter for DF Tau may be due to its status as an unresolved binary.)  In most stars there is good agreement between the excesses from the echelle and SpeX regimes, indicating that any systematic errors between the two approaches for finding $E_\lambda$ are not large. 

\addtocounter{figure}{-1}
\begin{figure*}
\epsscale{1}
\plotone{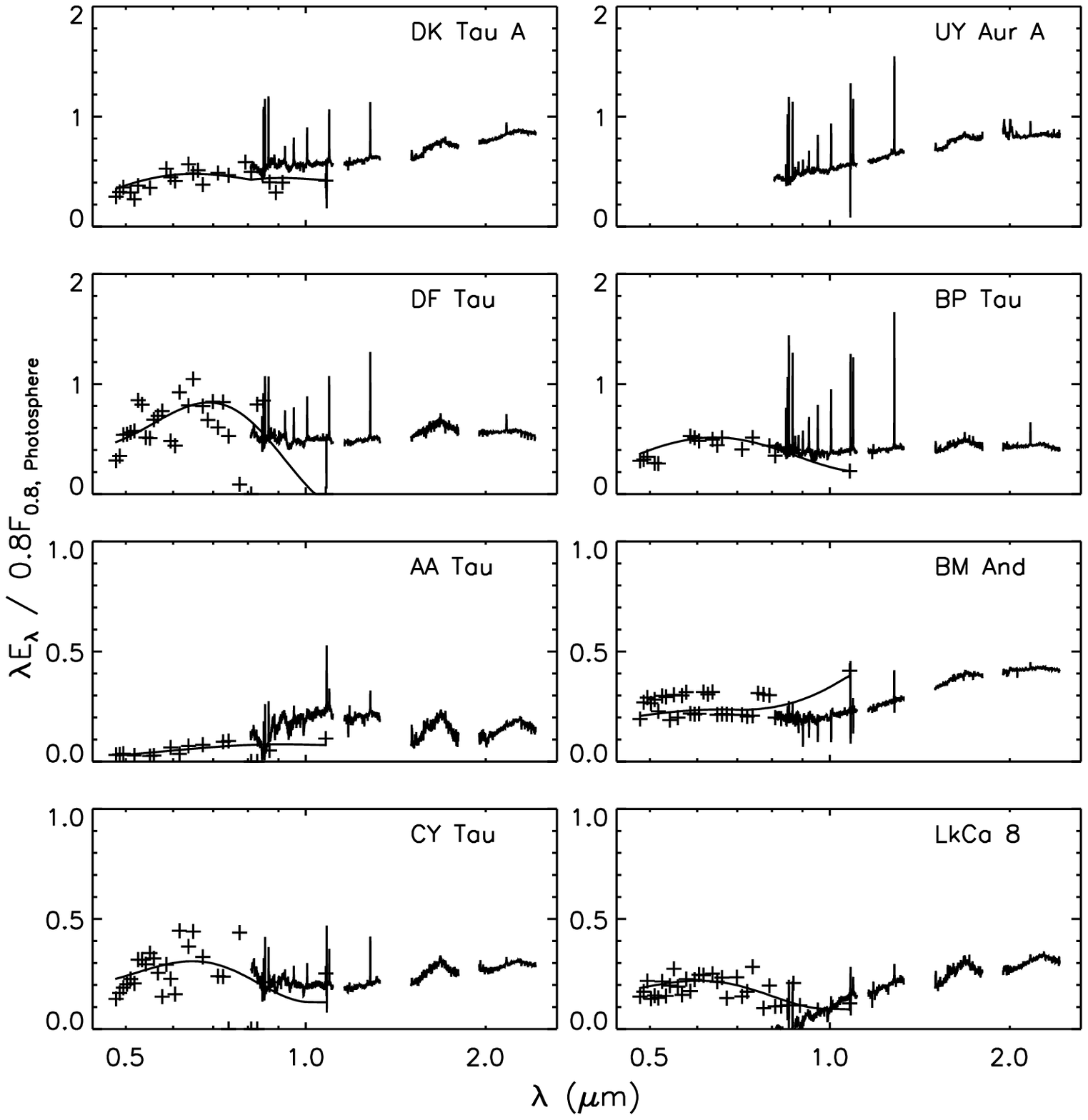}
\figcaption{cont.\label{f.excessb}}
\end{figure*}

To facilitate evaluation of the excess emission among the sample objects, we compute a median excess $E_{\Delta\lambda}$ in each of seven wavelength intervals $\Delta\lambda$.  The intervals, defined in Table~\ref{t.contex}, roughly follow the passbands of standard photometric filters $V$, $R$, $I$, $Y$, $J$, $H$, and $K$. The corresponding $E_{\Delta\lambda}$ in each wavelength interval are listed in Table~\ref{t.contex} for each star, where $E_V$ through $E_I$ are from the echelle spectra, and $E_Y$ through $E_K$ are from the SpeX spectra.  As for $\lambda E_{\lambda}$ in Figure~\ref{f.excessa}, the median excess $E_{\Delta\lambda}$ in each wavelength interval is in units of the photospheric flux at 0.8 \micron. These median values will be useful in characterizing the $IYJ$ excesses of CTTS that span several orders of magnitude in disk accretion rate.

\begin{deluxetable*}{lccccccccccc}
\tablecaption{Median Excess Fluxes, Ratios, and Slopes\label{t.contex}}
\tabletypesize{\footnotesize}
\setlength{\tabcolsep}{0.05in}
\tablewidth{0pt}
\tablehead{\colhead{} & \colhead{$E_V$} & \colhead{$E_R$} & \colhead{$E_I$} & \colhead{$E_Y$} & \colhead{$E_J$} & \colhead{$E_H$} & \colhead{$E_K$} & \colhead{} & \colhead{} & \colhead{} & \colhead{} \\
\colhead{} & \colhead{(0.50--} & \colhead{(0.60--} & \colhead{(0.70--} & \colhead{(0.95--} & \colhead{(1.16--} & \colhead{(1.50--} & \colhead{(1.95--} & \colhead{} & \colhead{} & \colhead{} & \colhead{} \\
\colhead{Object} & \colhead{0.60)} & \colhead{0.70)} & \colhead{0.80)} & \colhead{1.11)} & \colhead{1.33)} & \colhead{1.80)} & \colhead{2.40)} & \colhead{$E_Y/E_V$} & \colhead{$E_Y/E_K$} & \colhead{$S_J$} & \colhead{$S_K$}}
\startdata
AA Tau\dotfill   & 0.06 & 0.07 & 0.08 & 0.17 & 0.13 & 0.07 & 0.06 & 2.78 & 2.96 & \phs0.06 & \phs0.38 \\
AS 353A\dotfill  & 2.70 & 2.32 & 2.01 & 1.60 & 1.09 & 0.69 & 0.57 & 0.59 & 2.80 & $-$0.12  & \phs0.11 \\
BM And\dotfill   & 0.33 & 0.29 & 0.25 & 0.17 & 0.17 & 0.18 & 0.16 & 0.51 & 1.07 & \phs0.21 & \phs0.01 \\
BP Tau\dotfill   & 0.68 & 0.63 & 0.50 & 0.30 & 0.26 & 0.22 & 0.16 & 0.44 & 1.82 & \phs0.09 & $-$0.06  \\
CW Tau\dotfill   & 1.16 & 1.10 & 0.95 & 0.76 & 0.80 & 0.85 & 0.75 & 0.66 & 1.02 & \phs0.19 & \phs0.03 \\
CY Tau\dotfill   & 0.40 & 0.38 & 0.29 & 0.15 & 0.13 & 0.13 & 0.11 & 0.39 & 1.45 & \phs0.19 & \phs0.07 \\
DF Tau\dotfill   & 0.91 & 0.99 & 0.84 & 0.38 & 0.32 & 0.29 & 0.21 & 0.42 & 1.82 & \phs0.08 & $-$0.06  \\
DG Tau A\dotfill   & 1.03 & 0.87 & 0.84 & 0.53 & 0.41 & 0.37 & 0.30 & 0.52 & 1.75 & \phs0.04 & \phs0.06 \\
DK Tau A\dotfill   & 0.63 & 0.59 & 0.48 & 0.44 & 0.39 & 0.35 & 0.31 & 0.70 & 1.43 & \phs0.08 & \phs0.10 \\
DL Tau\tablenotemark{a}\dotfill   & (1.87) & (1.90) & (1.75) & 1.31 & 0.96 & 0.70 & 0.50 & 0.71 & 2.62 & $-$0.04  & $-$0.03  \\
DO Tau\dotfill   & 1.36 & 1.18 & 0.85 & 0.64 & 0.63 & 0.59 & 0.53 & 0.47 & 1.21 & \phs0.16 & \phs0.10 \\
DR Tau\dotfill   & 5.01 & 4.58 & 3.65 & 2.43 & 2.21 & 2.02 & 1.68 & 0.48 & 1.45 & \phs0.09 & \phs0.08 \\
HN Tau A\dotfill   & 2.25 & 1.73 & 1.27 & 0.68 & 0.49 & 0.50 & 0.53 & 0.30 & 1.28 & \phs0.06 & \phs0.25 \\
LkCa 8\dotfill   & 0.31 & 0.26 & 0.19 & 0.09 & 0.13 & 0.13 & 0.11 & 0.29 & 0.81 & \phs0.38 & \phs0.15 \\
RW Aur A\dotfill & \nod & \nod & \nod & 1.27 & 0.82 & 0.63 & 0.49 & \nod & 2.57 & $-$0.09  & \phs0.11 \\
UY Aur A\dotfill   & \nod & \nod & \nod & 0.42 & 0.41 & 0.38 & 0.30 & \nod & 1.38 & \phs0.15 & $-$0.05
\enddata
\tablecomments{$E_{\Delta\lambda}$ are the median excesses $F_{\lambda, \rm total}-F_{\lambda, \rm photosphere}$ over the wavelength ranges whose extents are indicated in microns, scaled so that $F_{\lambda, \rm photosphere}=1$ at 0.8 \micron.  $S_{\Delta\lambda}$ are slopes within the indicated bands; see \S\ 4.}
\tablenotetext{a}{For DL Tau, $E_V$, $E_R$, and $E_I$ are multiplied by 1.7 to correct for a shift between optical and IR data attributed to variability.}
\end{deluxetable*}

\section{BEHAVIOR OF THE EXCESS EMISSION SPECTRA}

\begin{figure*}
\plotone{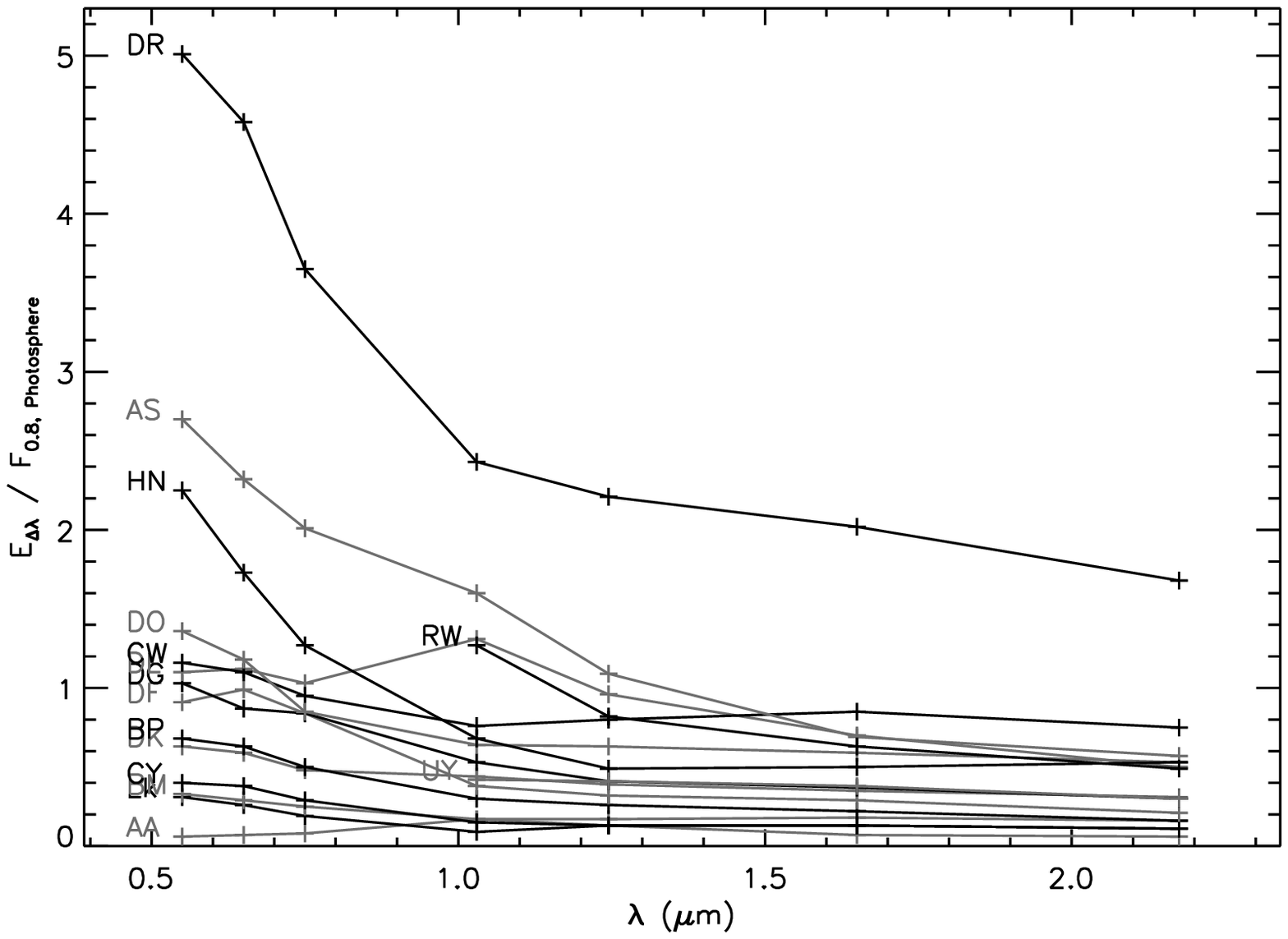}
\figcaption{Comparison of the median excesses $E_{\Delta \lambda}$ for all stars in our sample in wavelength intervals that approximate the bandpasses $V$, $R$, $I$, $Y$, $J$, $H$, and $K$ as defined in Table~\ref{t.contex}.  This simplified plot of the emission excesses from Fig.~\ref{f.excessa} allows a comparison across the full sample of CTTS.  All excesses are in units of the photospheric flux at 0.8 \micron.\label{f.elamall}}
\end{figure*}

\begin{figure*}
\plotone{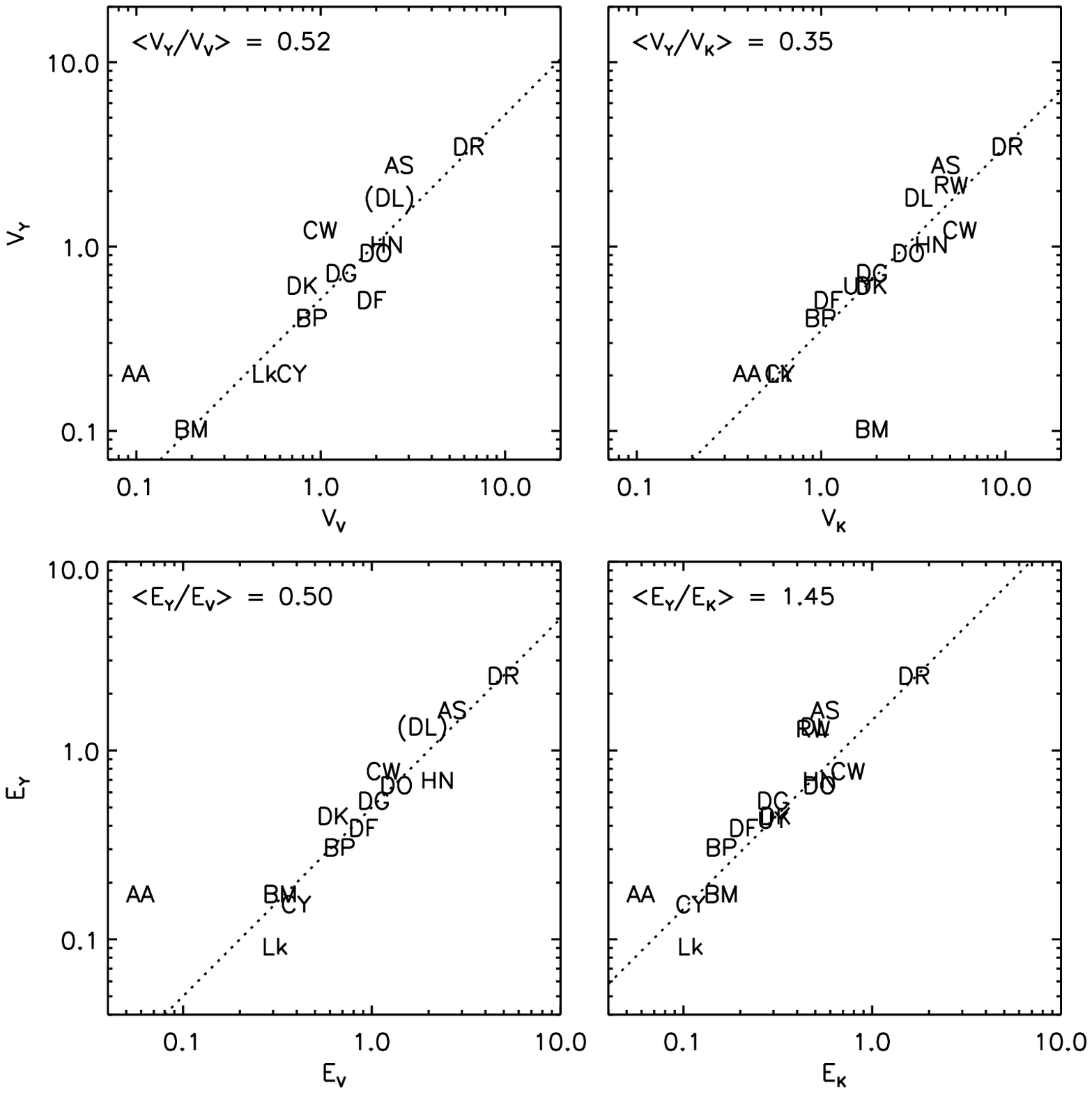}
\figcaption{Relations among excess emission at $V$, $Y$, and $K$ (Table~\ref{t.contex}) expressed both as veiling $V_{\Delta \lambda}$ and emission $E_{\Delta \lambda}$. {\em Top:} $V_Y$ versus $V_V$ ({\em left}) and $V_K$ ({\em right}).  {\em Bottom:} $E_{\Delta \lambda}$ for the same pairs of wavelengths in units of the star's photospheric flux at 0.8~\micron. Dotted lines correspond to the median ratios labeled in each panel. To account for variability, the $V$ excess of DL Tau has been increased by the ratio of its SpeX to its echelle excess in the region where they overlap.\label{f.contex}}
\end{figure*}

The CTTS in our sample show excess emission over the full range of our wavelength coverage. Although the median excesses in the seven wavelength bands approximating $V$, $R$, $I$, $Y$, $J$, $H$, and $K$ oversimplify the full SEDs, a plot of the $E_{\Delta\lambda}$ sequence for all 16 stars in the sample, shown in Figure~\ref{f.elamall}, is instructive.  The overall shape of the excess emission is similar among the full sample, where, in units of the photospheric flux at 0.8~\micron, the excess declines from $V$ through $Y$ to $K$ while covering a range in magnitude at $Y$ from three times the photospheric flux (DR Tau) to only a few tenths of the photospheric flux (AA Tau, LkCa~8).  This can be shown quantitatively by plotting the median excesses at 1 \micron, $E_Y$, against the median excesses at the shortest and longest wavelength intervals in our dataset, $E_V$ and $E_K$. Comparing the median excesses in these three bands is particularly instructive, since $E_V$ will include a contribution from the blue excess used to model disk accretion rates, $E_K$ will include a contribution from the near-infrared excess attributed to sublimating dust in the disk, and $E_Y$ will have minimal contributions from these two components. This comparison is shown in Figure~\ref{f.contex}, both for the excesses $E_V$, $E_Y$, and $E_K$ and for the corresponding continuum veilings $V_V$, $V_Y$, and $V_K$.  The ratios of these excesses are remarkably consistent from star to star when plotted logarithmically.  The median values of the excess ratios are $E_Y/E_V=0.50$ and $E_Y/E_K=1.45$, and the median values of the veiling ratios are $V_Y/V_V=0.52$ and $V_Y/V_K=0.35$. The figure and columns 9 and 10 of Table~\ref{t.contex} show that the dispersions of the excess ratios for individual stars around these median ratios are not large; the interquartile ranges are 0.42 to 0.66 for $E_Y/E_V$ and 1.25 to 2.20 for $E_Y/E_K$.

\begin{figure*}
\plotone{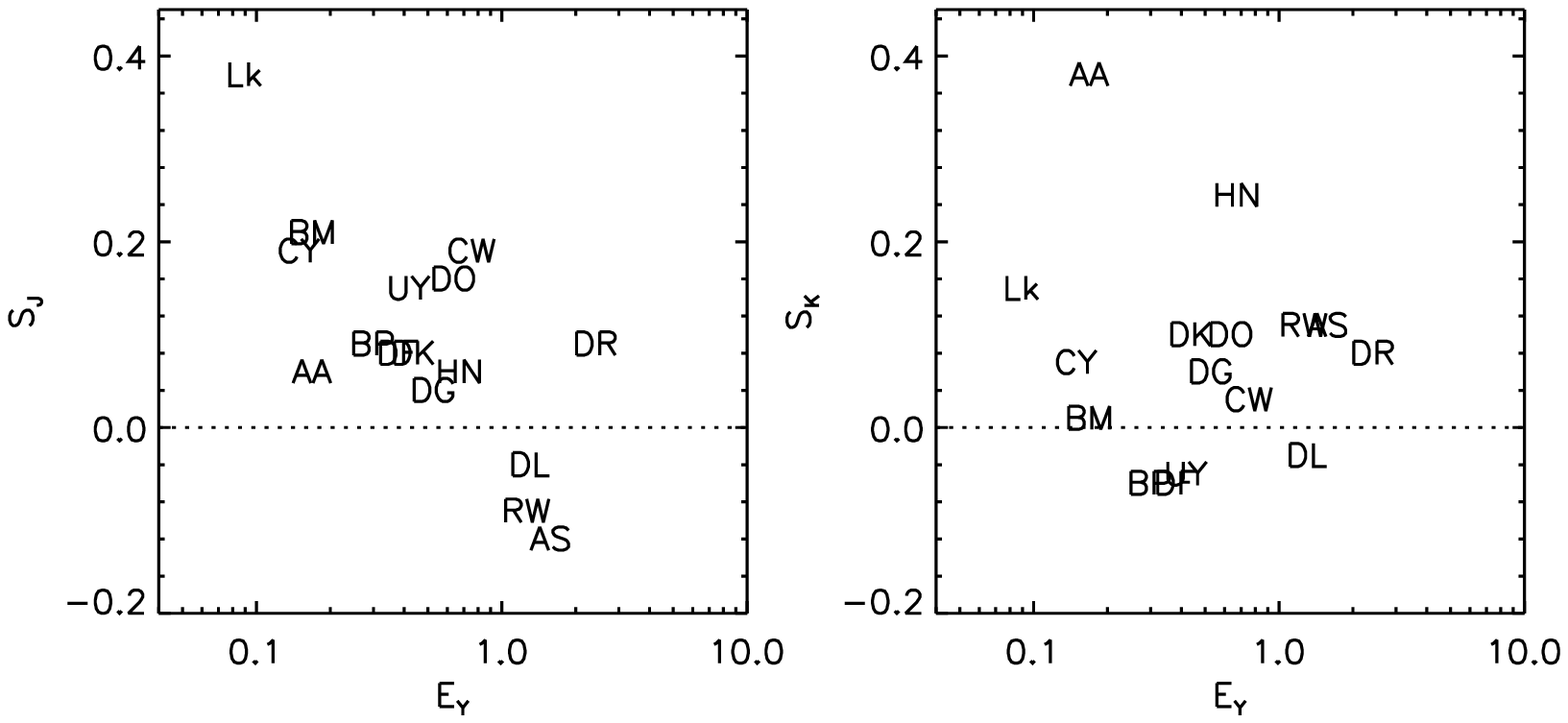}
\figcaption{Slopes of $\lambda E_\lambda$ in the $J$ band ({\em left}) and the $K$ band ({\em right}) versus the normalized flux of the continuum excess in $Y$.  The slope across a band $\Delta \lambda$ is defined as  $S_{\Delta \lambda}=(\lambda_f E_{\lambda_f}-\lambda_i E_{\lambda_i})/\lambda_i E_{\lambda_i}$, where $\lambda_i$ is the shortest wavelength and $\lambda_f$ is the longest wavelength in the band.\label{f.slopes}}
\end{figure*}

Despite the correlations in $E_V$, $E_Y$, and $E_K$ shown in Figure~\ref{f.contex}, the $\lambda E_\lambda$ curves of individual stars in Figure~\ref{f.excessa} have different slopes in particular wavelength intervals that are masked when we take an average across each band.  We illustrate this in Figure~\ref{f.slopes} for the $J$ and $K$ bands, since the spectra there are usually well described by a single slope. The slopes, included in Table~\ref{t.contex}, are defined as $S_{\Delta \lambda}=\left(\lambda_f E_{\lambda_f}-\lambda_i E_{\lambda_i}\right)/\lambda_i E_{\lambda_i}$, where $\lambda_i$ and $\lambda_f$ are, respectively, the shortest and longest wavelengths in the regions defined in Table~\ref{t.contex}. The slopes in both bands run roughly from $-0.1$ to 0.25 but show no correlation with each other or with the magnitude of $E_Y$. The most notable outliers are three stars with large $E_Y$ and negative $S_J$: DL Tau, RW Aur A, and AS 353 A.  Looking back at Figure~\ref{f.excessa},  these three stars are distinctive, with $\lambda E_\lambda$ rising steeply to shorter wavelengths from the $J$ through the $I$ band. In contrast, the other stars with large $E_Y$ are much flatter through the $I$ band and have excesses that rise to longer wavelengths through the $J$ band, with slopes from 0.1 (DR Tau, HN Tau A, and DG Tau A) to 0.2 (DO Tau and CW Tau).

The two wavelength regions sensitive to surface gravity also show a range of behaviors in $\lambda E_\lambda$ (see \S\ 3.4 and Fig.~\ref{f.compare}). In the $H$ band, a hump in $\lambda E_\lambda$ becomes more prominent in stars with small excesses, as might be expected from a surface gravity mismatch. However, since a rise to shorter wavelengths between 0.80 and 0.85 \micron, which would also be an expected from a surface gravity mismatch, is not seen in these stars, the structure in the $H$ band may be inherent in the excess.  

\begin{figure}
\resizebox{\hsize}{!}{\plotone{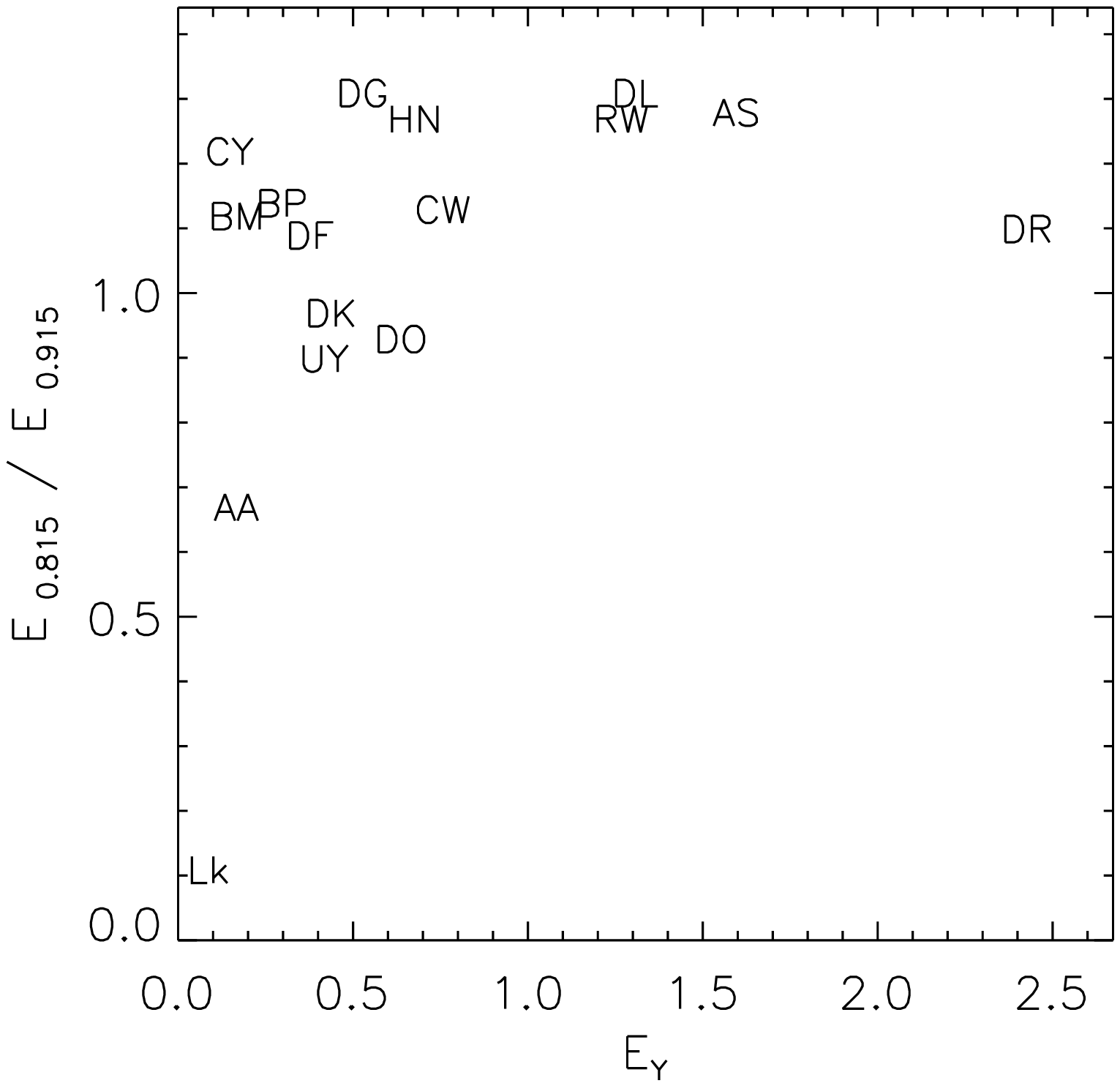}}
\figcaption{Ratios of excess flux on the blue side of the Paschen jump (0.815 \micron) to that on the red side of the Paschen jump (0.915~\micron) versus $E_Y$.\label{f.pajump}}
\end{figure}

Only a few stars have excess spectra that rise to shorter wavelengths between 0.80 and 0.85 \micron\ (e.g., DL Tau).  While this behavior may be related to the Paschen jump at 0.8207~\micron, no star shows a distinct Paschen discontinuity in $E_\lambda$.  Figure~\ref{f.pajump} plots an index measuring the Paschen jump (the ratio of the excess emission at 0.815~\micron\ to that at 0.915 \micron) against $E_Y$.  In contrast to the Balmer jump, which is more pronounced for stars with weaker blue excess \citep{her08}, there is no correlation between the Paschen jump and the $IYJ$ excess.

\begin{figure*}
\plotone{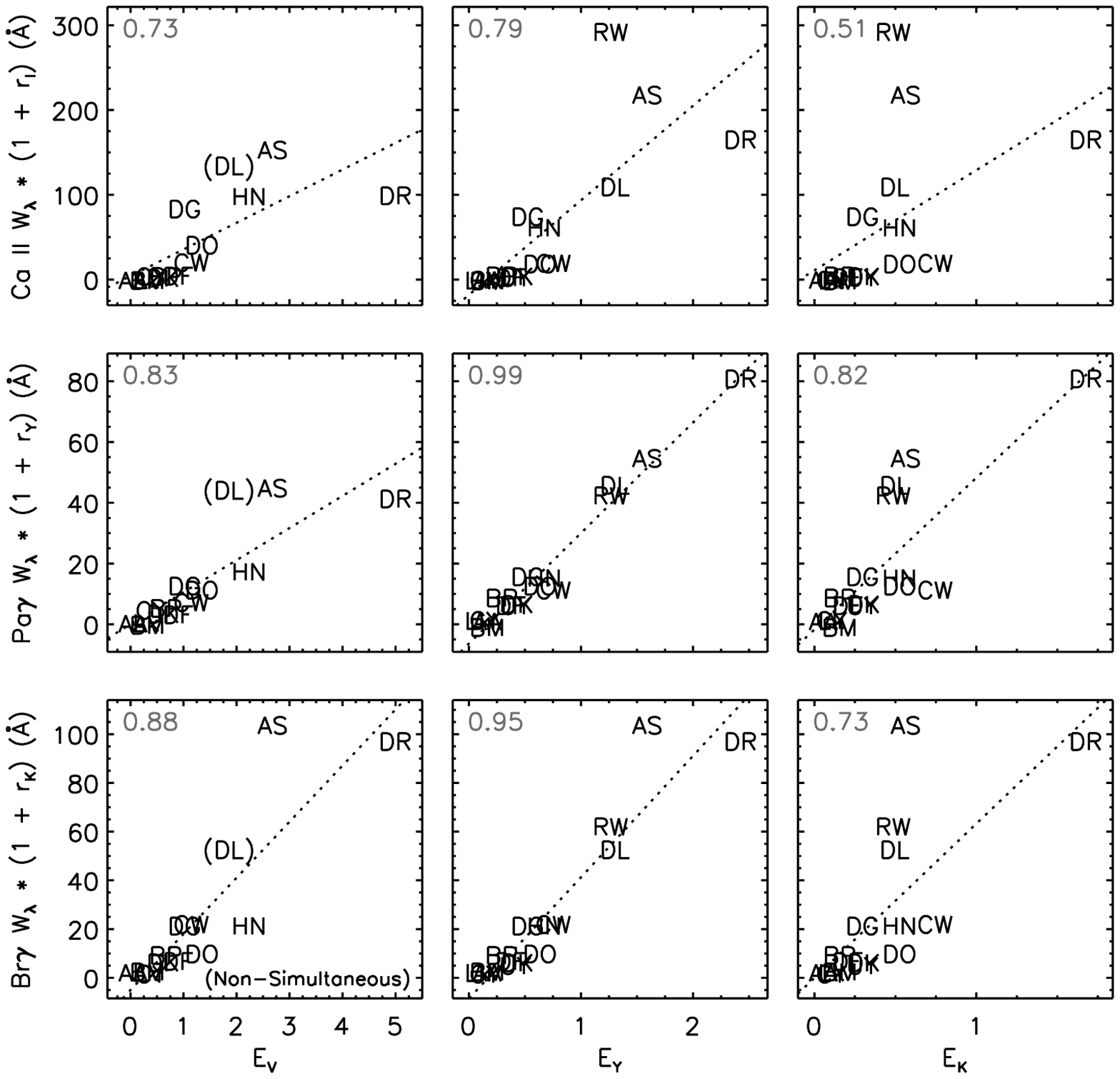}
\figcaption{Comparison of veiling-corrected equivalent widths of three strong emission lines to the median excesses $E_V$, $E_Y$, and $E_K$ (Table~\ref{t.contex}). The correlation coefficients appear in the upper left of each panel.  All line/continuum pairs are simultaneous, either from SpeX or echelle data, except for Br$\gamma$ (SpeX) and $E_V$ (echelle).  As in Fig.~\ref{f.contex}, $E_V$ for DL Tau is adjusted to account for variability.\label{f.linex}}
\end{figure*}

We next compare the average excess continuum emission in the three wavelength intervals $V$, $Y$, and $K$ to the fluxes in strong emission lines that have been shown to correlate with non-simultaneously determined disk accretion rates \citep{muz98a,muz98b}.  In Figure~\ref{f.linex} we compare the excess continuum fluxes $E_V$, $E_Y$, and $E_K$ to veiling-corrected equivalent widths for the three emission lines \ion{Ca}{2} $\lambda8500$, Pa$\gamma$, and Br$\gamma$.  The directly measured equivalent widths (see Table~\ref{t.linex}) are multiplied by the factor $1+r_\lambda$ to make them relative to the photospheric flux, thus facilitating comparisons of stars with different levels of veiling.  Although fluxes in each of the three lines correlate with the median excesses, the correlations show much more scatter for $E_K$ than for $E_V$ or $E_Y$.  The linear correlation coefficients, listed in the upper left corner of each panel, are lowest for \ion{Ca}{2} $\lambda8500$ in all three wavelength intervals, but this arises from a break in the dependence of the \ion{Ca}{2} $\lambda8500$ flux on the excess strength, which is weak at low excess and stronger at high excess, such that a single linear fit does not represent the situation adequately. As will be discussed in a forthcoming paper examining the full complement of emission lines in our combined HIRES, NIRSPEC, and SpeX data set, this break in the \ion{Ca}{2} versus $E_{\Delta \lambda}$ relation corresponds to a transition from the line being dominated by a narrow component to being dominated by a broad component. In contrast, the correspondence of the hydrogen lines, which always show a broad component \citep{edw06}, to $E_V$ or $E_Y$ is essentially linear, although the better correlation with $E_Y$ suggests that the $Y$ band excess may be more closely affiliated with the region responsible for the hydrogen line emission.

In summary, we find that the levels of the excess emission in three wavelength intervals that correspond to the beginning, middle, and end of our spectral coverage ($E_V$, $E_Y$, and $E_K$) are well correlated with each other and with emission line fluxes recognized as accretion diagnostics. Furthermore, while the shape of the excess over the full wavelength range is not strictly uniform from star to star (e.g., slopes vary over restricted wavelength intervals), it is roughly similar from star to star, making it unlikely that random effects such as large cool spots or undetected companions are major influences on the derivation of the excess.

\section{MODELING THE {\em IYJ} EXCESS EMISSION}

In this section we present an initial characterization of the temperature, size, and luminosity of the region(s) responsible for the excess emission between 0.48 and 2.4~\micron.  According to previous investigations of the optical/UV and near-infrared excesses, the excess in this wavelength regime will have a contribution at its shortest wavelengths from hot accretion spots on the star \citep{cal98} and contributions at its longest wavelengths from the dust sublimation zone in the disk \citep{muz03}. Based on these studies, we adopt $T_{\rm hot}=8000$~K and $T_{\rm cool}=1400$~K for these two regions.  The magnitudes of their contributions will then depend on the area of each region projected along the line of sight. For both regions, we express this area as a filling factor $f$ relative to the projected area of the star, which can be less than one (smaller than the star) or greater than one (larger than the star).  Typical filling factors for hot accretion spots are less than 1\% in most cases but can be up to 5\% in a few of the most active accretors \citep{cal98,gul00}. If a zone in the disk is treated as a face-on, flat annulus, its filling factor relative to the star has a simple expression.  In this case, $f=2R_0\Delta R$, where $R_0$ is the distance from the origin to the center of the annulus, and $\Delta R$ is the width of the annulus, both in units of the stellar radius.  For example, an annulus of width $1~R_*$ at a distance of $10~R_*$ has $f=20$, i.e., its area is 20 times the projected area of a star with radius $1~R_*$.  \citet{muz03} modeled SpeX 2--5 \micron\ spectra of nine CTTS and found that their excess emission over this wavelength range was well described by $T\sim1400$~K dust from sublimation radii in the disk  between 7 and 32 $R_*$.  With this range of  $R_0$ and assuming $\Delta R\sim0.1~R_0$,  the corresponding filling factors for the 1400~K component range from $\sim10$ to 300.

\subsection{Constraints from $E_V$, $E_Y$, and $E_K$}

\begin{figure*}
\includegraphics[angle=90,width=\hsize]{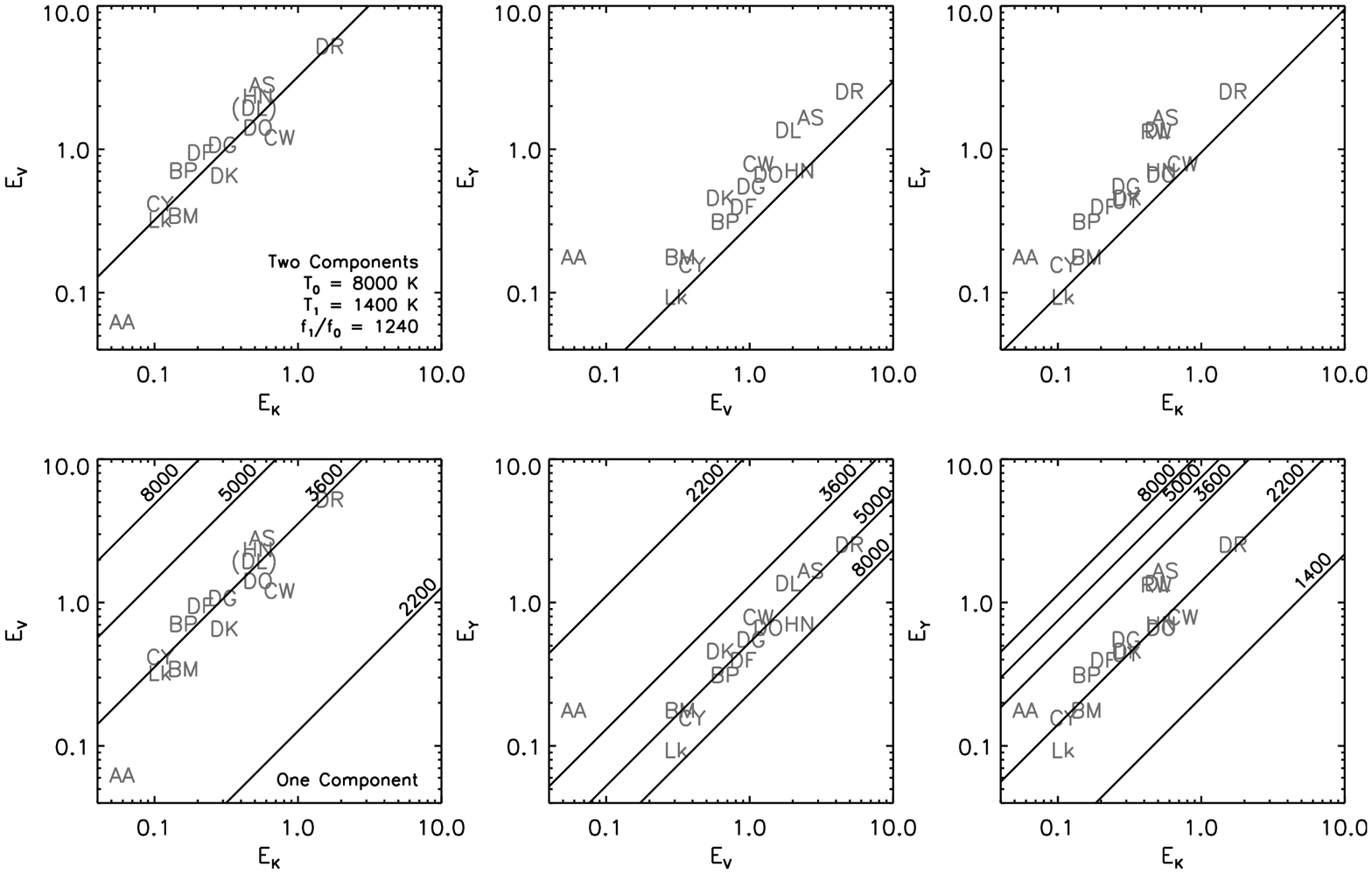}
\figcaption{Comparison of observed median excesses $E_V$, $E_Y$, and $E_K$ to those of blackbodies. {\em Top row:}  Excesses from two components with temperatures 8000 K and 1400 K and a ratio of filling factors of 1240. While this two-component fit can describe $E_V/E_K$, the corresponding ratios $E_Y/E_V$ and $E_Y/E_K$ are smaller than observed by more than a factor of two. {\em Bottom row:} Correspondence of single-temperature blackbodies to excess ratios at different pairs of wavelengths.  All excesses are in units of the star's photospheric flux at 0.8 \micron.\label{f.modelex}}
\end{figure*}

To illustrate the inadequacy of only two regions with $T_{\rm hot}=8000$~K and $T_{\rm cool}=1400$~K to account for the observed $IYJ$ excess, we first look at the median excesses at the extremes and center of our wavelength coverage, $E_V$, $E_Y$, and $E_K$ (Fig.~\ref{f.contex}), before turning to the full shape of the excess. In the upper row of Figure~\ref{f.modelex}, we compare the observed excess ratios at pairs of wavelengths to those from a two-component model with blackbody temperatures $T_{\rm hot}=8000$~K and  $T_{\rm cool}=1400$~K. The observed median relation for $E_V/E_K$ is well described by a combination of these two temperature regimes if the filling factor of the cool component is 1240 times that of the hot component.  However, the required filling factors for the hot region fall between 0.015 (LkCa 8) and 0.22 (DR Tau), in contrast to typical values of 0.001--0.05 inferred from accretion shock models \citep{cal98}.  Furthermore, the observed ratios $E_Y/E_V$ and $E_Y/E_K$ are on average twice as large as the best two-component model predicts.

The unexpectedly large $E_Y$ measurements suggest that a third component, with a temperature intermediate to an accretion spot with $T_{\rm hot}=8000$~K and sublimating dust with $T_{\rm cool}=1400$~K, must contribute to the observed excess.  We can roughly constrain the allowed range of temperatures $T_{\rm int}$ for this region by comparing the $E_V$, $E_Y$, and $E_K$ relations to flux ratios from a single-temperature thermal radiator as shown in the lower row of Figure~\ref{f.modelex}. We see that $E_Y/E_V$ is too large for 8000 K, corresponding instead to a thermal radiator of 5000 K, and $E_Y/E_K$ is far too large for 1400 K, corresponding instead to a thermal radiator of 2200 K.  We infer from this that the temperature(s) of the third component must lie between 2200 and 5000 K. 

We explore whether a model with three blackbody components, characterized by three temperatures and three filling factors, can explain the observed excesses. In order to constrain parameter space, we specify two temperatures from the outset,  $T_{\rm hot}=8000$~K and $T_{\rm cool}=1400$~K, and search for combinations of $T_{\rm int}$, $f_{\rm hot}$, $f_{\rm int}$, and $f_{\rm cool}$ that fulfill the observed $E_Y/E_V$ and $E_Y/E_K$ relations.  If, for example, we specify $T_{\rm int}$, then unique ratios of filling factors are required among the three temperature regions.  This is true because the ratio of excess fluxes at two wavelengths can be written as 
\begin{eqnarray}
\lefteqn{\frac{E_{\lambda_1}}{E_{\lambda_2}}=} \\
 & & \frac{B_{\lambda_1}(T_{\rm hot})+B_{\lambda_1}(T_{\rm int})f_{\rm int}/f_{\rm hot}+B_{\lambda_1}(T_{\rm cool})f_{\rm cool}/f_{\rm hot}}{B_{\lambda_2}(T_{\rm hot})+B_{\lambda_2}(T_{\rm int})f_{\rm int}/f_{\rm hot}+B_{\lambda_2}(T_{\rm cool})f_{\rm cool}/f_{\rm hot}}, \nonumber
\end{eqnarray} where $B_{\lambda_i}(T_j)$ is the Planck function for temperature $T_j$ evaluated at wavelength $\lambda_i$.  A system of two such equations (e.g., for $E_Y/E_V$ and $E_Y/E_K$), can then be solved to find expressions for $f_{\rm int}/f_{\rm hot}$ and $f_{\rm int}/f_{\rm cool}$ that depend only on the three assumed temperatures and the observed median excess ratios.  

\begin{figure}
\resizebox{\hsize}{!}{\plotone{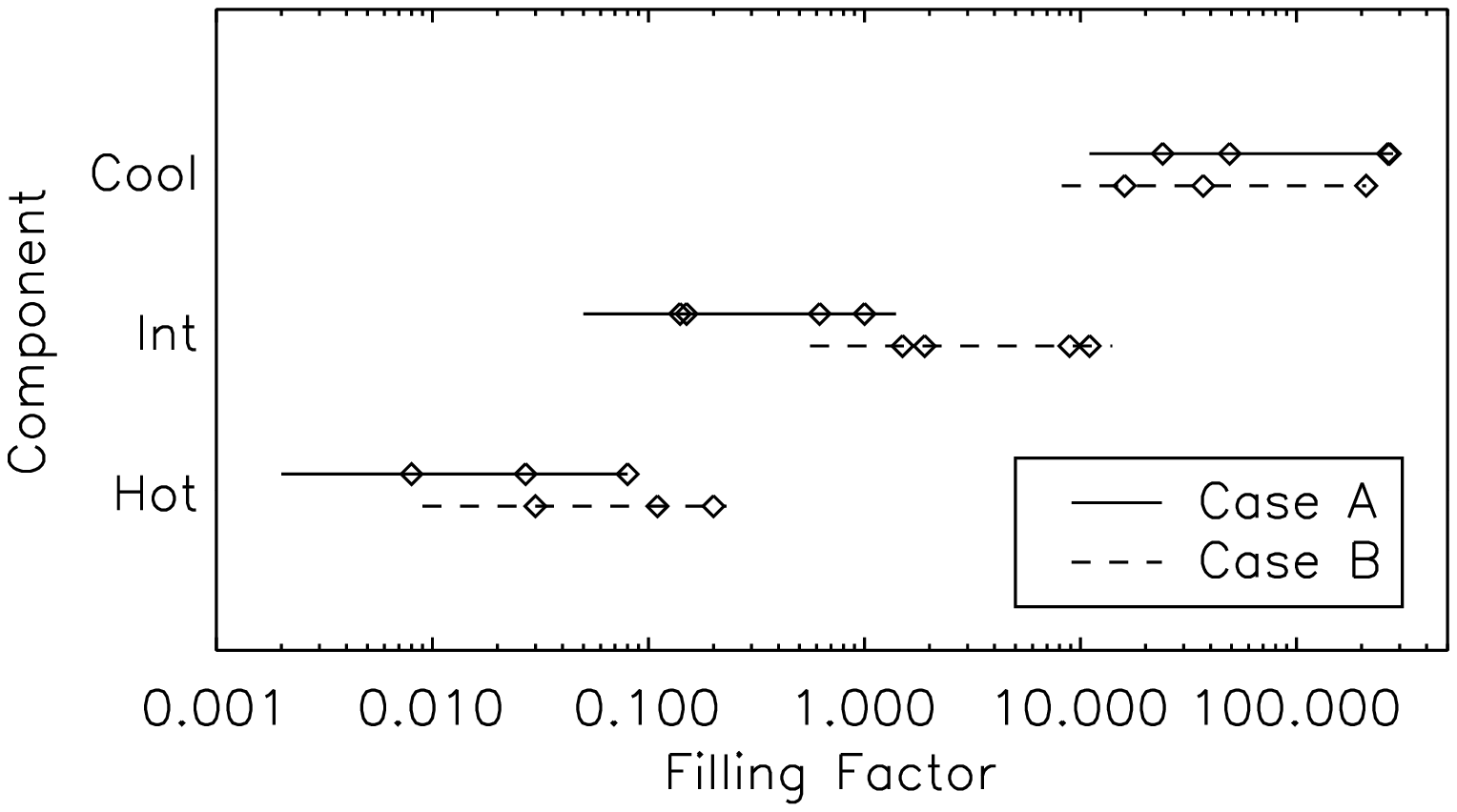}}
\figcaption{Filling factors, in units of the projected area of the star, required to describe the observed excesses in a three-component model.  The solid (Case A; $T_{\rm int}=5000$ K) and dashed (Case B; $T_{\rm int}=2500$ K) lines show the ranges for all stars (Table~\ref{t.ranges}). Diamonds mark the individual filling factors for the four stars with SEDs modeled in \S\ 5.2 (Fig.~\ref{f.triplefit}).\label{f.fillf}}
\end{figure}

We choose two values for $T_{\rm int}$ as an illustration.  In Case A, $T_{\rm int}=5000$~K, as might correspond to lower-temperature annuli surrounding the hot accretion spots on the stellar surface. In Case B, $T_{\rm int}=2500$~K, as might correspond to disk gas inside the dust sublimation radius.  Given the median observed ratios $E_Y/E_V=0.50$ and $E_Y/E_K=1.45$, the ratios of filling factors will be $f_{\rm hot}/f_{\rm int}/f_{\rm cool}=1/30/6100$ for Case A and $f_{\rm hot}/f_{\rm int}/f_{\rm cool}=1/62/910$ for Case B.  Although the filling factors in these models remain in fixed ratio to one another, the magnitudes of the observed excesses require that the magnitudes of the filling factors span a range of about a factor of 25. The resulting values if the photospheric temperature is 4000 K are illustrated in Figure~\ref{f.fillf} and are listed in Table~\ref{t.ranges}.  The implied sizes for the hot and cool components are reasonable for the physical regimes attributed to them in the above scenarios. In both cases, the hot-spot filling factors are mostly $\ll1$ with a maximum of 0.23, and the cool dust filling factors range from 8 to 280, in line with the crude assumption of a dust sublimation region at a distance $R_0$ between 7 and 32 $R_*$ from the star and with width $\sim0.1~R_0$.  The corresponding size range for the intermediate-temperature component in Case A ($T_{\rm int}=5000$ K) is $f_{\rm int}=0.05$ to 1.4 and in Case B ($T_{\rm int}=2500 $K) is $f_{\rm int}$ = 0.6 to 14. For the low-excess stars, if $T_{\rm int}\sim5000$ K, then $f_{\rm int}$ would be small enough to correspond to warm annuli around accretion hot spots on the stellar surface, but for the high-excess stars, this is an unphysical interpretation since $f_{\rm int}$ can be $>1$. In contrast, if $T_{\rm int}\sim2500$ K, then $f_{\rm int}$ is always large in comparison to the star, but smaller than a ring at the dust sublimation radius.

While a single blackbody of $T_{\rm int}$ with size in exact proportion to that of accretion spots on the photosphere is likely an oversimplified description of the origin of the $IYJ$ excess for all the CTTS in our sample, we infer that all three components have sizes that scale with the magnitude of the excess.  There is already evidence for this in models of accretion shocks, where filling factors up to 5\% are invoked to explain the largest blue continuum excesses \citep{gul00}, and in the derived dust sublimation radii, which are largest for CTTS with high disk accretion rates \citep{eis10}. It is surprising that, for temperatures between 2500 and 5000 K, the requisite size of the intermediate-temperature component is comparable to or exceeds the stellar surface area for those stars with large $IYJ$ excesses.  In the next subsection we go beyond the simple median ratios of $E_V$, $E_Y$, and $E_K$, and we experiment with fitting the full SED of the excess emission.

\begin{deluxetable*}{lcccccc}
\tablecaption{Filling Factors Yielding Median Excess Ratios for Fixed $T_{\rm int}$\label{t.ranges}}
\tabletypesize{\footnotesize}
\tablewidth{0pt}
\tablehead{\colhead{Case} & \colhead{$f_{\rm int}/f_{\rm hot}$} & \colhead{$f_{\rm cool}/f_{\rm hot}$} & \colhead{$f_{\rm cool}/f_{\rm int}$} & \colhead{Range of $f_{\rm hot}$} & \colhead{Range of $f_{\rm int}$} & \colhead{Range of $f_{\rm cool}$}}
\startdata
A: $T_{\rm int}=5000$~K & 30 & 6100 & 200 & 0.002--0.05 & 0.05--1.4 & 11--280 \\
B: $T_{\rm int}=2500$~K & 62 & 910 & 15 &  0.009--0.23 & 0.56--14 & 8.2--210
\enddata
\tablenotemark{\phn}{}
\end{deluxetable*}

\subsection{SED Fitting}

Given the need for at least three temperature components with sizes that increase with the magnitude of the excess to account for the median ratios among $E_V$, $E_Y$, and $E_K$, we attempt here to reproduce the full shape of the excess emission $\lambda E_\lambda$. We continue to use the Case A and Case B scenarios described in the previous section, with $T_{\rm hot}=8000$~K, $T_{\rm cool}=1400$~K, and $T_{\rm int}=5000$~K (Case A) or 2500 K (Case B).

\begin{figure*}
\plotone{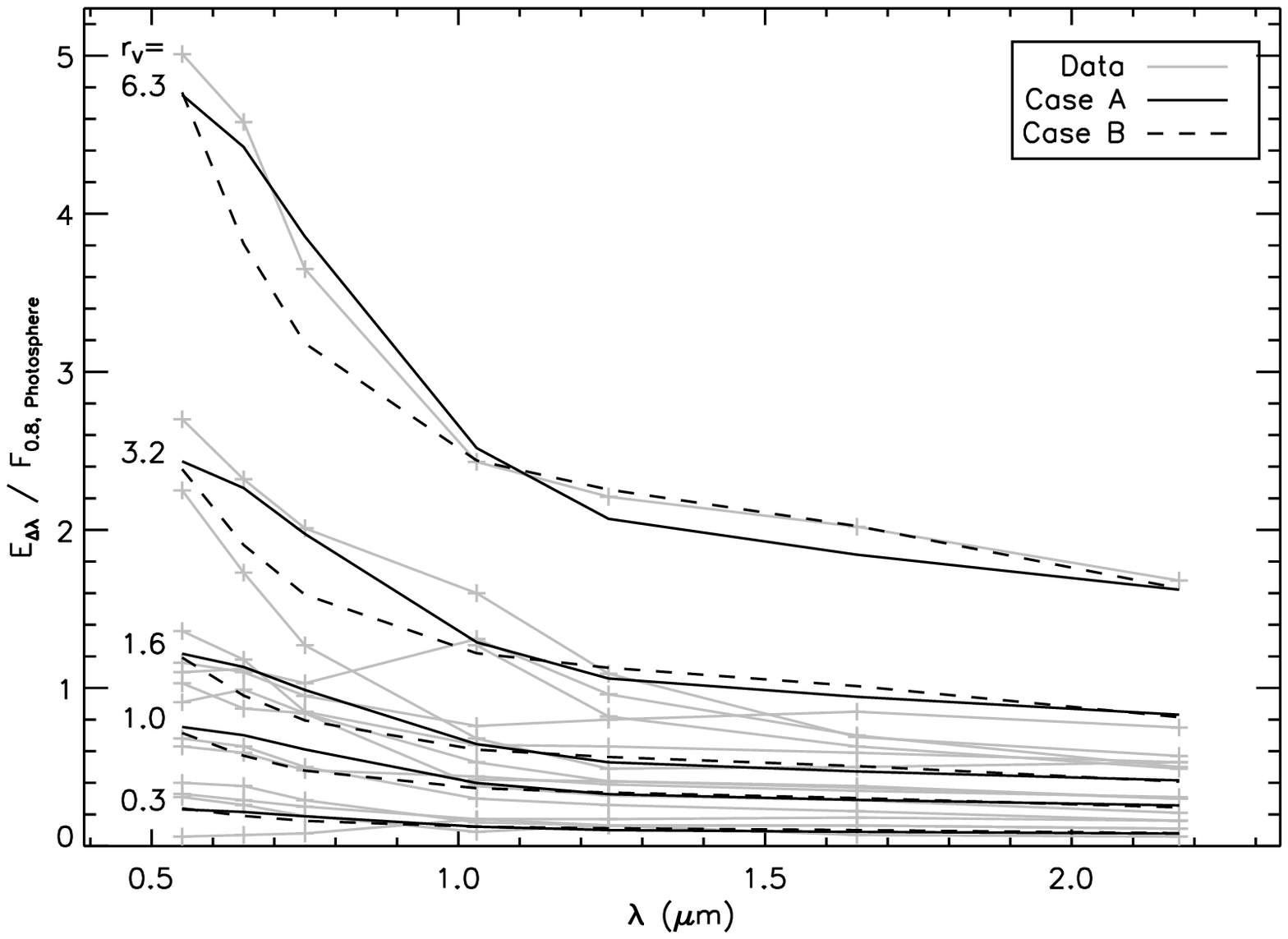}
\figcaption{Comparison of $E_{\Delta \lambda}$ for Case A ($T_{\rm int}=5000$ K) and Case B ($T_{\rm int}=2500$ K) to observations (gray lines, see Fig.~\ref{f.elamall} for star identification). The strengths of the model excesses are characterized by the corresponding veiling at the shortest wavelength of our dataset, $r_V=0.3$ to 6.3.  Both three-component fits are reasonable approximations to the observed spectra, although Case A is a better fit at short wavelengths and Case B is a better fit at long wavelengths.\label{f.elamab}}
\end{figure*}

A comparison of calculated spectra from Case A and Case B to the median excesses at seven wavelength intervals approximating the $V$, $R$, $I$, $Y$, $J$, $H$, and $K$ bandpasses (Table~\ref{t.contex} and Fig.~\ref{f.elamall}) is shown in Figure~\ref{f.elamab}.  Either scenario accounts reasonably well for the general behavior of the observed excesses, but $T_{\rm int}=5000$ K is a better match to the excesses between $V$ and $J$, and $T_{\rm int}=2500$~K is a better match between $J$ and $K$.

\begin{figure*}
\plotone{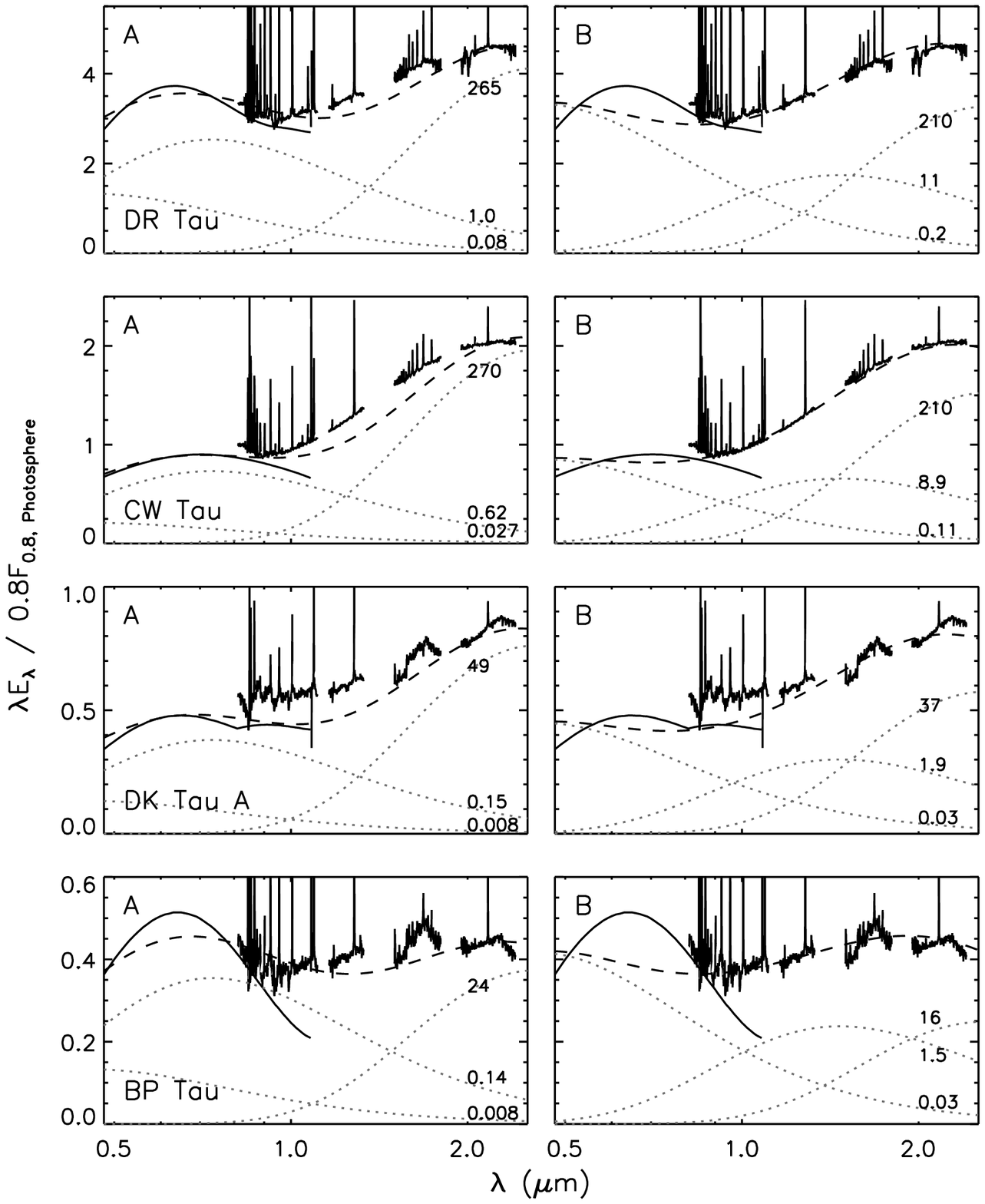}
\figcaption{Comparison of three-component blackbody fits for Case A ($T_{\rm int}=5000$ K; {\em left}) and Case B ($T_{\rm int}=2500$ K; {\em right}) to the emission excess $\lambda E_\lambda$ (solid lines, from Fig.~\ref{f.excessa}) for four stars.  Dotted lines show the three components of each model, and dashed lines show their sum.  The required filling factors for each component are labeled (also see Table~\ref{t.fits}).\label{f.triplefit}}
\end{figure*}

Further comparison of Cases A and B to the observed excesses can be made by dropping the requirement of a uniform ratio of filling factors for all stars and letting the filling factors for each region match the relations among $E_V$, $E_Y$, and $E_K$ for each individual star, which can differ by factors of several from the median ratios.  We show this in Figure~\ref{f.triplefit} for four representative sources: two high-excess stars (DR Tau and CW Tau) and two moderate-excess stars (DK Tau A and BP Tau).  We compare their full observed $\lambda E_{\lambda}$ spectra derived from the echelle and SpeX data to both Case A and Case B models.  The filling factors chosen for each star are listed in Table~\ref{t.fits} and marked in Figure~\ref{f.fillf}.  In addition to the summed fit, we also show the individual contribution from each component. Again we see that, although these simple three-component models give reasonable fits to $\lambda E_{\lambda}$, the warmer Case A better represents the shorter-wavelength excess, while the cooler Case B better represents the $J$ through $K$ bands.  We conclude that, although Cases A and B both describe the observed excesses reasonably well, the source of the excess emission in this wavelength regime is more likely to arise from regions with a range of temperatures between 2200 and 5000 K.

\begin{deluxetable*}{lccccccc}
\tablecaption{Three-Component Fit Parameters for Selected CTTS\label{t.fits}}
\tabletypesize{\footnotesize}
\tablewidth{0pt}
\tablehead{\colhead{Object} & \colhead{Case} & \colhead{$f_{\rm int}/f_{\rm hot}$} & \colhead{$f_{\rm cool}/f_{\rm hot}$} & \colhead{$f_{\rm cool}/f_{\rm int}$} & \colhead{$f_{\rm hot}$} & \colhead{$f_{\rm int}$} & \colhead{$f_{\rm cool}$}}
\startdata
DR Tau\dotfill & A & 12.5 & 3310  & 265  & 0.080 & 1.00 & 265 \\
CW Tau\dotfill & A & 23.0 & 10000 & 435  & 0.027 & 0.62 & 270 \\
DK Tau A\dotfill & A & 18.8 & 6130  & 327  & 0.008 & 0.15 & 49 \\
BP Tau\dotfill & A & 17.5 & 3000  & 171  & 0.008 & 0.14 & 24
\vspace{2pt}\\
\hline\vspace{-5pt}\\
DR Tau\dotfill & B & 55.0 & 1050  & 19.1 & 0.20  & 11.0 & 210 \\
CW Tau\dotfill & B & 80.9 & 1910  & 23.6 & 0.11  & 8.9  & 210 \\
DK Tau A\dotfill & B & 70.3 & 1370  & 19.5 & 0.03  & 1.9  & 37 \\
BP Tau\dotfill & B & 60.0 & 640   & 10.7 & 0.03  & 1.5  & 16
\enddata
\tablenotemark{\phn}{}
\end{deluxetable*}

\subsection{Residual Spectrum and Luminosity\\of the {\em IYJ} Excess}

\begin{figure*}
\plotone{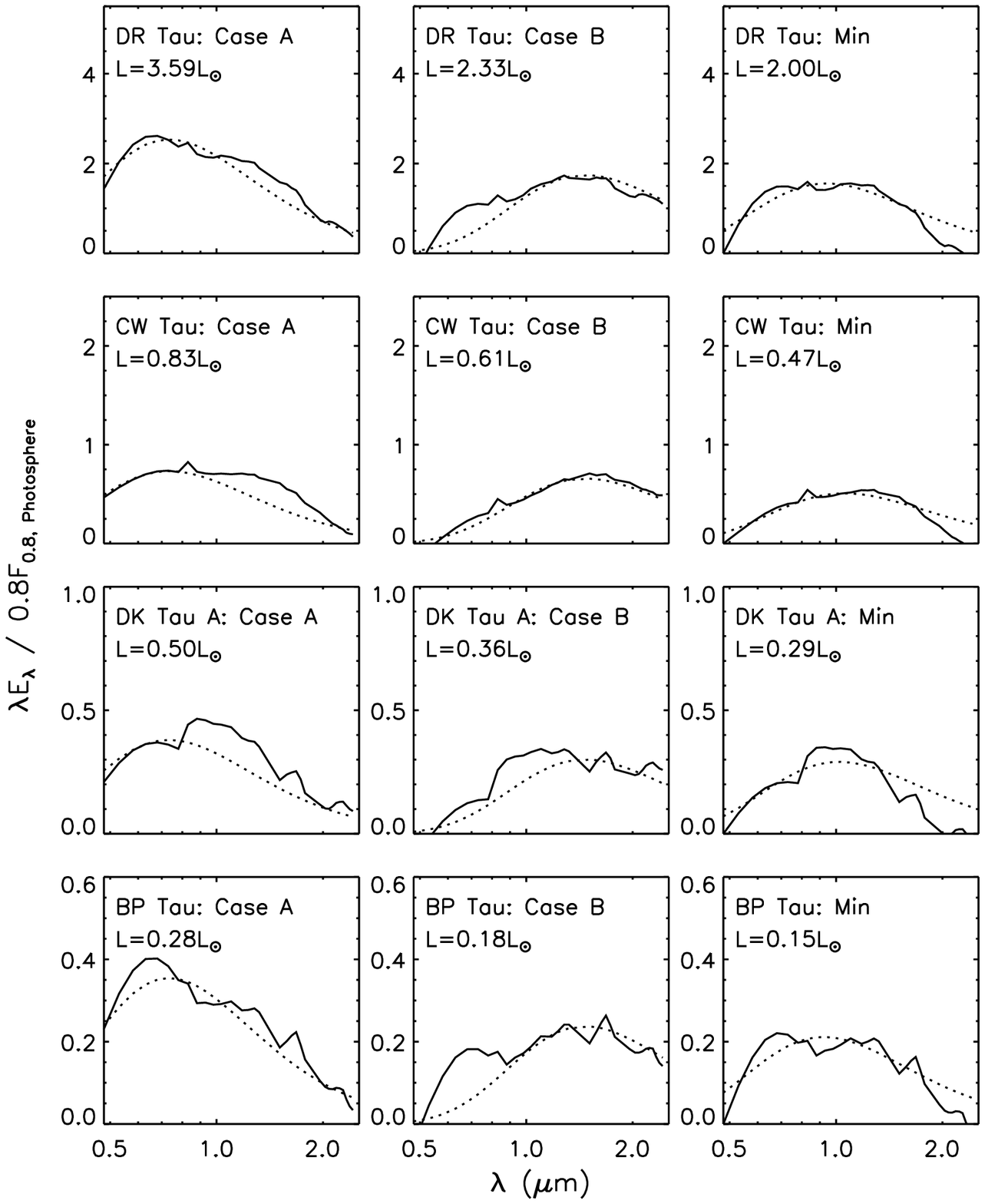}
\figcaption{Residual $\lambda E_\lambda$ for the intermediate-temperature component with three different assumptions for four representative stars. The Case A ({\em left}) and Case B  ({\em center}) residuals are derived by subtracting the hot and cool excess contributions in Fig.~\ref{f.triplefit} from the observed excess. Minimum-luminosity residuals are in the right column, for the extreme assumption that the excess at 0.48 \micron\ is entirely from the hot component and at 2.4 \micron\ is entirely from the cool component (see Table~\ref{t.lum}). Dotted lines correspond to blackbodies of $T_{\rm int}=5000$ K ({\em left}), $T_{\rm int}=2500$ K ({\em center}), and best fits ({\em right}), where the latter are between 3500 and 4000 K.  
\label{f.exlum}}
\end{figure*}

To uniquely determine the excess emission spectrum that arises solely from the region of intermediate temperature, one must specify the contributions from the hot and cool components, which would require even greater wavelength coverage than we present here.  We can, however, constrain the spectral shape and luminosity of the intermediate-temperature component by estimating the contributions from the hot and cool components and subtracting these contributions to generate a residual excess.  We show such residual spectra in the first two columns of Figure~\ref{f.exlum} for the four stars whose excesses were fit with Case A and B scenarios in Figure~\ref{f.triplefit}. For clarity, only SpeX data were used where they overlap with echelle data, and the $\lambda E_\lambda$ curves were median-filtered and rebinned to a sparse wavelength grid in order to remove emission lines, placing the focus on the overall continuum shape.  For each star, the residual (solid line) is compared to the blackbody (dotted line) for the assumed $T_{\rm int}$ of Case A or B. We find good agreement between the residual spectrum and the corresponding blackbody.  

An alternate approach is to define a minimum residual emission by assigning all of the excess at 0.48 \micron\ to an accretion spot with a blackbody spectrum of $T_{\rm hot}=8000$~K, assigning all of the excess at 2.4 \micron\ to dust with a blackbody spectrum of $T_{\rm cool}=1400$~K, and then removing their contributions to the observed $\lambda E_{\lambda}$ between 0.48 and 2.4 \micron.  By comparison, in Case A the flux at 0.48 \micron\ has a substantial contribution from both the intermediate and hot components, and in Case B the flux at 2.4 \micron\ has a substantial contribution from both the intermediate and cool components (see Fig.~\ref{f.triplefit}).  These minimum residuals are shown in the last column of Figure~\ref{f.exlum} (solid lines) and are compared to blackbodies (dotted lines). The blackbody spectra are in good agreement with the minimum residuals, with temperatures intermediate to Cases A and B: 3900 K for DR Tau, 3470 K for CW Tau, 3600 K for DK Tau, and 3980 K for BP Tau.  

\begin{figure}
\plotone{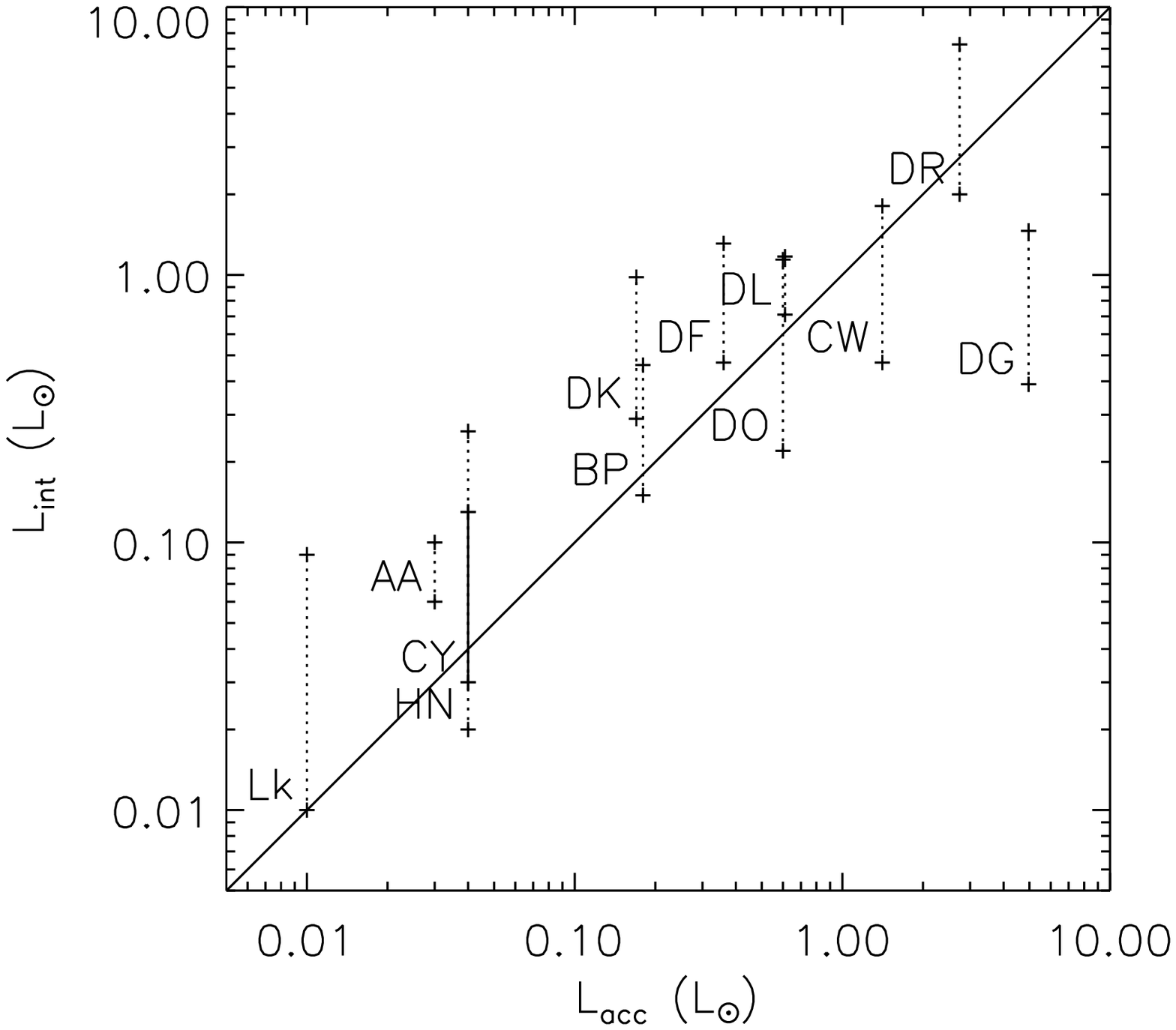}
\figcaption{Comparison of minimum and maximum estimates for the luminosity of the intermediate-temperature component giving rise to the $IYJ$ excess for each star to previously published accretion luminosities. The maximum assumes that all of the excess between 0.48 and 2.4 \micron\ comes from the intermediate-temperature component, and the minimum assumes that all of the excess at 0.48 \micron\ comes from an 8000 K accretion shock, while all of the excess at 2.4 \micron\ comes from 1400 K warm dust (also see Table~\ref{t.lum}).\label{f.complum}}
\end{figure}

The luminosities corresponding to these three approaches for evaluating the residual emission from the intermediate-temperature component are shown in each panel of Figure~\ref{f.exlum}. Luminosities are calculated by numerical integration of the excess emission spectra, initially expressed in units of $L_*$ as ratios of excess luminosities to photospheric luminosities.  To convert to $L_\sun$, these are multiplied by literature estimates of the stellar luminosities (Table~\ref{t.lum}). For each of the four stars, the range in luminosities from the three approaches is less than a factor of two. It is obviously of interest to compare the estimated luminosities of the intermediate-temperature components to previously published accretion luminosities. For all 14 CTTS in our sample that have both echelle and SpeX spectra, we do this by estimating both the minimum and maximum luminosity of the intermediate component from the observed excess. For the minimum luminosity we use the approach just described for all stars, forcing the intermediate-temperature contribution at the shortest and longest wavelengths to zero. For the maximum luminosity we attribute the entire excess between 0.48 and 2.4 \micron\ to the intermediate-temperature component.  Again, the luminosities are first found in photospheric units  and then converted to physical units using published values of the photospheric luminosities. The minimum and maximum estimates, along with the stellar and accretion luminosities from the literature, are listed for each star in Table~\ref{t.lum} and compared graphically in Figure~\ref{f.complum}.  The minimum luminosities run from 0.01 to 2.4 $L_\sun$ (0.03 to 1.2 $L_*$), while the maximum luminosities run from 0.09 to 7.8 $L_\sun $ (0.22 to 4.2 $L_*$). Literature accretion and photospheric luminosities were taken from \citet{gul98a} or \citet{cal98} whenever possible (all but two stars) in order to ensure a uniform approach.  Figure~\ref{f.complum} shows that the minimum and maximum luminosity estimates of the $IYJ$ excess bracket the accretion luminosity determined from the blue excess, suggesting that the actual luminosity of this component is comparable to the previously derived accretion luminosity for each star. The implication is that total accretion luminosities for most CTTS are a factor of two higher than previously estimated from modeling the blue excess emission.

\begin{deluxetable*}{lccccccc}
\tablecaption{Minimum and Maximum Excess Luminosities\label{t.lum}}
\tabletypesize{\footnotesize}
\tablewidth{0pt}
\tablehead{\colhead{Object} & \colhead{$L_*(L_\sun)$} & \colhead{$L_{\rm acc}(L_\sun)$} & \colhead{Refs.\tablenotemark{a}} & \colhead{$L_{\rm min}(L_*)$\tablenotemark{b}} & \colhead{$L_{\rm min}(L_\sun)$} & \colhead{$L_{\rm max}(L_*)$\tablenotemark{c}} & \colhead{$L_{\rm max}(L_\sun)$}}
\startdata
AA Tau\dotfill & 0.71 & 0.03 & 2,2 & 0.08 & 0.06 & 0.14 & 0.10 \\
AS 353A\dotfill & 3.72 & \nod & 3 & 0.66 & 2.44 & 2.10 & 7.82 \\
BM And\dotfill & 7.24 & \nod & 5 & 0.07 & 0.50 & 0.31 & 2.28 \\
BP Tau\dotfill & 0.93 & 0.18 & 2,2 & 0.16 & 0.15 & 0.49 & 0.46 \\
CW Tau\dotfill & 1.35 & 1.41 & 4,1 & 0.35 & 0.47 & 1.34 & 1.81 \\
CY Tau\dotfill & 0.46 & 0.04 & 2,2 & 0.06 & 0.03 & 0.28 & 0.13 \\
DF Tau\dotfill & 1.97 & 0.36 & 2,2 & 0.24 & 0.47 & 0.67 & 1.31 \\
DG Tau A\dotfill & 1.74 & 4.96 & 3,1 & 0.23 & 0.39 & 0.84 & 1.46 \\
DK Tau A\dotfill & 1.45 & 0.17 & 2,2 & 0.20 & 0.29 & 0.68 & 0.98 \\
DL Tau\dotfill & 0.68 & 0.61 & 3,1 & 0.75 & 0.71 & 1.72 & 1.17 \\
DO Tau\dotfill & 1.01 & 0.60 & 2,2 & 0.22 & 0.22 & 1.13 & 1.14 \\
DR Tau\dotfill & 1.74 & 2.74 & 3,1 & 1.15 & 2.00 & 4.17 & 7.26 \\
HN Tau A\dotfill & 0.19 & 0.04 & 2,2 & 0.12 & 0.02 & 1.37 & 0.26 \\
LkCa 8\dotfill & 0.41 & 0.01 & 2,2 & 0.03 & 0.01 & 0.22 & 0.09
\enddata
\tablenotetext{a}{References for stellar and accretion luminosities.}
\tablenotetext{b}{Intermediate-component luminosity if the excess emission at 0.48~\micron\ is due entirely to hot spots and the excess emission at 2.4~\micron\ is due entirely to warm dust.}
\tablenotetext{c}{Total excess luminosity from 0.48 to 2.4 \micron.}
\tablerefs{(1) \citealt{cal98}; (2) \citealt{gul98a}; (3) \citealt{har95}; (4) \citealt{ken95}; (5) \citealt{ros99}.}
\end{deluxetable*}

In summary, from our very simple modeling of the excess emission between 0.48 and 2.4~\micron, we find that three temperature components whose filling factors increase with increasing total excess account reasonably well for the observations.  The temperature structure of the intermediate component is not well specified, but gas with one or more temperatures between 2200 and 5000~K is likely.  It seems inescapable that for stars with the largest $IYJ$ excesses (highest disk accretion rates), the requisite filling factor is comparable to the surface area of the star if $T_{\rm int}$ is 5000 K and is up to an order of magnitude higher if $T_{\rm int}$ is only 2500~K. This is unreasonably large to attribute to warm annuli around spots in the shock-heated photosphere and points to an origin in the disk, accretion flow, or wind. The luminosity from this region is comparable to previous estimates of the accretion luminosity.

\section{DISCUSSION}

We have demonstrated the presence of an excess continuum in accreting T Tauri stars between 0.48 and 2.4 \micron\ with an intensity that scales with other indicators of disk accretion, such as the blue excess continuum and atomic emission lines.  The similarity in the gross spectral characteristics of the $IYJ$ excess among a diverse sample of stars makes it unlikely that contamination from random factors, such as large cool spots or undetected companions, contributes much to the derived excess. We discuss some implications of the $IYJ$ excess, which has a temperature between that attributed to $\sim8000$ K hot spots in the shock-heated photosphere (the dominant contributor to the excess shortward of 0.5 \micron; \citealt{cal98}) and a raised rim of $\sim1400$ K dust at the dust sublimation radius in the disk (the dominant contributor to the excess from 2 to 4.8 \micron; \citealt{muz03}). Our simple modeling suggests that the intermediate component has a temperature between 2200 K and 5000 K, and it likely has a range of temperatures between these limits. One issue to consider is the effect of this newly recognized contribution to CTTS excess continuum emission on the determination of disk accretion rates. Another is the absence of any spectral region where broadband colors sample the young stellar photosphere, which makes extinction determinations from broadband colors difficult.  Most exciting is the potential of identifying a heretofore unrecognized aspect of the star-disk interaction in accreting systems.

\vspace{0.5in}

\subsection{Effect of the {\em IYJ} Excess on Derived\\Disk Accretion Rates}

Although we have identified a new source of continuum emission between  0.48 and 2.4 \micron, we cannot uniquely determine its properties due to spectral overlap with other known sources of excess emission.  However, reasonable upper and lower limits indicate the excess luminosity in this wavelength range is comparable to previously determined accretion luminosities. Provided the excess is not simply reprocessing radiation from accretion hot spots, this indicates that a significant amount of accretion energy is being dissipated in the region we call the intermediate-temperature component. It is not immediately obvious what the implication is for estimates of disk accretion rates. When disk accretion rates are determined from spectrophotometry blueward of 0.5 \micron, this emission is well represented by accretion shock models  \citep{cal98} where the underlying assumption is that the excess luminosity derives from kinetic energy liberated by material free-falling to the stellar surface in accretion columns. Thus accretion rates determined under these assumptions should be reasonable estimates of the amount of matter actually landing on the stellar surface if the $IYJ$ excess arises from a region other than the accretion shock. However, if some of the $IYJ$ excess arises from annuli of warm gas surrounding a central, small accretion hot spot at the magnetic footpoint, then accretion rates will need to be revised upward accordingly. 

More troubling is the commonly used technique of estimating disk accretion rates from large samples of CTTS based on line-veiling measurements in the Paschen continuum longward of 0.5 \micron. Since this approach relies on a bolometric correction appropriate for a hot accretion shock, it will overestimate accretion luminosities, progressively more so at longer wavelengths, since much of the excess will come from the intermediate-temperature component. In order to correctly determine CTTS accretion rates based on line veilings, the total accretion luminosity from all temperature components would first need to be found from absolutely simultaneous spectrophotometric data extending from shortward of the Balmer jump in the $U$ band to longward of the gas continuum excess, probably out to the $K$ band.  Evaluation of reddening with a good grid of both subdwarf and dwarf standards through this full wavelength region would be necessary to understand the tendency to find larger $A_V$ at longer wavelengths in both WTTS and CTTS (see next section) and thus to appropriately correct the observed emission for extinction. Then to convert line-veiling measurements to accretion luminosities, the relationship between veiling observed over a small wavelength interval and the total accretion luminosity must be established in order to know what kind of bolometric correction to apply.  Finally, a model that includes all sources of excess luminosity would need to be constructed so that accretion luminosities can be turned into disk accretion rates. 

At present, we can conclude only that the excess $IYJ$ emission has a luminosity that is comparable to previously derived accretion luminosities and that accretion rates based on line veiling longward of 0.5 \micron\ are suspect.   An additional problem is that, for many CTTS, photospheric properties such as luminosity, radius, and mass will need to be re-evaluated in light of the fact that much of the flux between 0.5 and 1 \micron\ previously attributed to the photosphere is actually excess emission. Flux-calibrated spectrophotometry from the UV to the near IR in combination with sophisticated models of all contributors to the excess emission will be needed to improve our understanding of accretion luminosities and disk accretion rates. 

\subsection{Effect of the {\em IYJ} Excess on Extinction Estimates}

\begin{figure*}
\epsscale{0.975}
\plotone{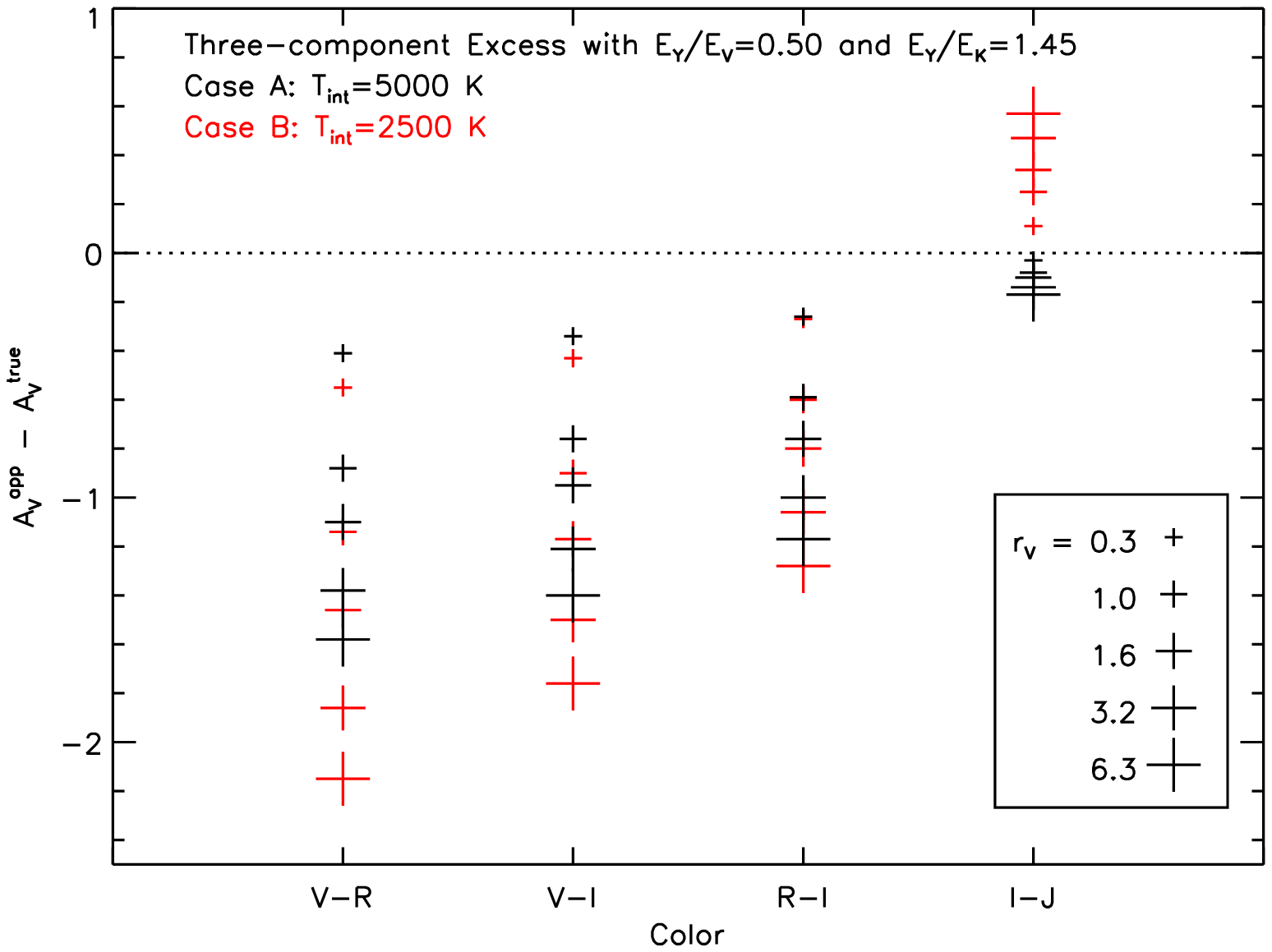}
\figcaption{The difference between the apparent reddening ${A_V}^{\rm app}$  and the true reddening ${A_V}^{\rm true}$ as a function of color if observed CTTS colors are assumed to be photospheric. The underlying photosphere is 4000 K and the excess continuum is the sum of three components corresponding to Cases A (black) and B (red). The magnitude of the excess is identified by the veiling at $V$, with $r_V$ from 0.3 to 6.3, corresponding to the SEDs shown in Fig.~\ref{f.elamab}.\label{f.av}}
\epsscale{1}
\end{figure*}

The demonstration that CTTS have excess continuum emission in the region intermediate to the well studied blue and infrared excesses, including the $I$, $Y$, and $J$ bands, makes it clear that there are no wavelength regions where photometric colors sample the bare photosphere. Thus, using observed colors to determine $A_V$ is subject to error.  Many compilations of extinctions for CTTS use red colors to derive $A_V$, such as $V-R$ \citep{ken95}, $V-I$ \citep{whi01}, $R-I$ \citep{eir02}, or $I-J$ \citep{mcc10}. To assess the effect of the $IYJ$ excess on determining $A_V$ from observed colors, we synthesized SEDs from 0.5 to 2 \micron\ that are composed of a 4000 K stellar photosphere with varying amounts of continuum excess and subjected to intervening extinction.  The continuum excess consists of three temperature components corresponding to the Case A and Case B of the previous section, with fixed $T_{\rm hot}$ and $T_{\rm cool}$, and $T_{\rm int}=2500$ or 5000 K. The magnitude of the excess is traced by the veiling at $V$, with $r_V$ ranging from 0.3 to 6.3. The resulting difference between the apparent and true $A_V$ as a function of color is shown in Figure~\ref{f.av}.  A failure to correctly account for the $IYJ$ excess emission leads to an underestimate of $A_V$ from all colors except $I-J$, which are often not observed simultaneously.  Although there is excess emission at $I$ and $J$, the intermediate components in Cases A and B have temperatures similar to the star, and thus the $I-J$ color, which primarily samples the intermediate component, happens to be nearly photospheric.  On the other hand, the underestimate of $A_V$ for heavily veiled stars can be as large as 2 magnitudes when using $V-R$ and 1 magnitude when using $R-I$. (The difference between the true and derived $A_V$ is equal to the difference between the $V$ magnitude of the bare photosphere and the $V$ magnitude of the excess-plus-photosphere system; thus, it depends only on the properties of the excess and is independent of the true $A_V$.)

An additional issue regarding extinction measurements in CTTS is the puzzling tendency to find $A_V$ from 0.8 to 2.4 \micron\ spectra that consistently exceed previous estimates. This could result from a number of effects that may all be operating in concert, including incorrect assessment of $A_V$ from red colors due to a large $IYJ$ excess, differences in surface gravity or temperature between the CTTS and dwarf spectral templates, and distortion of the spectrum by large cool spots or undetected close companions.  (We consider the last two options unlikely because of the similarity in the gross characteristics of the $IYJ$ excess in our 16 CTTS.)  This effect may be related to the unexplained behavior in the SEDs of some WTTS, which have increasingly anomalous colors for their spectral types toward longer wavelengths \citep{gul98a,gul98b}.  We tested the SpeX spectrum of the WTTS V819 Tau to see if this behavior could be explained by an $IYJ$ excess that arises from a blackbody source with $T=2500$, 3500, or 5000 K and is weak enough to produce negligible veiling shortward of 0.6 \micron. We find that such a weak excess does generate a higher $A_V$ with increasing wavelength, but only by a few tenths of a magnitude, much smaller than observed.  We reiterate that whatever gives rise to the large $A_V$ determined from the SpeX spectra does not significantly affect our characterization of the $IYJ$ excess, as described in \S\ 4, but it does indicate that extinctions to CTTS and WTTS need to be revisited in a more thorough way than can be done here. 

\subsection{A New Component in Accretion Disk Systems?}

Identifying the source of the $IYJ$ excess continuum requires that we define its SED in order to determine its temperature, density, opacity, and filling factor. Since at the shortest wavelengths in our data there is a contribution to the excess from the hot accretion spots, and at the longest wavelengths there is a contribution from the warm dust rim in the disk, we cannot extract the residual SED arising solely from the additional red emission.  It does appear to be broad and relatively smooth, as would be expected from blackbody radiation with a temperature (or temperatures) between 2200 and 5000~K.  Possible sources for the red continuum emission are warm annuli around accretion hotspots, gas in the disk inside the dust sublimation radius, and gas in the accretion flow or wind. The presence of multiple temperature components in the shock-heated photosphere would not be surprising; in fact, this is predicted by 3D magnetohydrodynamic numerical simulations of accretion in the presence of dipolar and more complex magnetic fields \citep{rom04,lon08}. This model may apply to stars with small $IYJ$ excesses, provided that about half of the accretion luminosity is radiated from these cooler regions. However, for stars with large $IYJ$ excesses, the required filling factors exceed the size of the star, making this an improbable scenario. The areas of the accretion flow and wind are larger than the surface area of the star, making these plausible locations for the $IYJ$ excess, although the requisite temperature is much cooler than the 10,000 K that is invoked for accretion streams as a source of hydrogen line emission \citep{muz01}.

The possibility that the $IYJ$ excess is due to dust-free gas in the disk inward of the raised rim at the dust sublimation radius deserves consideration. Theoretical considerations suggest that the dust-free inner gaseous disk should be optically thin with strong molecular emission \citep{muz04}, but this conclusion is called into question by \citet{dul10}. These authors note that there appears to be a deficit of molecules in the dust-free inner disk, which only rarely shows CO overtone or H$_2$O emission \citep{naj07}, and they suggest that molecules are destroyed in the inner gaseous disk, possibly from proximity to photodissociating radiation. They also present evidence that the inner gaseous disk may be optically thicker in CTTS than in intermediate mass Herbig Ae/Be stars, such as the behavior of linear polarization across the H$\alpha$ line, which appears to be scattered off a rotating inner gaseous disk \citep{vin05}.

Additional evidence for disk gas inside the dust sublimation radius comes from near-infrared spectro-interferometry in the $K$ band. Both spectroscopy and near-infrared interferometry indicate that the dust sublimation zone is at a radial distance in the disk of $\sim0.1$~AU for CTTS \citep{muz03,eis05,ake05,mil07}. Near-infrared spectro-interferometry at a resolution of $R=230$ across the $K$ band by \citealt{eis09} (and \citealt{eis07} \& \citealt{tan08a} for higher-mass AeBe stars) shows that the angular size of the near-infrared emission decreases with decreasing wavelength, indicating that the ring of warm dust is accompanied by warm, presumably gaseous material within the dust sublimation radius. ÊFor higher-mass stars than we have investigated here, the possibility that this excess may arise from non-blackbody small dust grains has also been considered \citep{ben10a,ben10b}. Detailed modeling (e.g., by \citealt{tan08b}) indicates that for gas opacity primarily due to free-free H and H$^-$ emission, the gas is partially optically thin with continuum emission primarily from material hotter than 2000--3000 K but cooler than 5000 K.  It is possible that the excess continuum in our broader spectral coverage comes at least in part from this gaseous region in the inner disk.

The inner gaseous disk, lying between the dust sublimation zone and the truncation of the disk by the stellar magnetosphere, is not yet well characterized but is a likely place for accretion energy to be dissipated. The gas that accretes onto the star must pass through this region, and magneto-centrifugual disk winds could be launched here. ÊIf the $IYJ$ continuum emission arises at least in part in this region, line emission would also be expected. Although the region appears to be depleted of molecules \citep{dul10}, some of the abundant atomic emission lines in CTTS spectra may arise here. Near-infrared spectro-interferometry shows that the Brackett emission in CTTS is unresolved \citep{eis10}, requiring formation inside the resolved dust sublimation region, and in this study we found a tight correlation between the Ê$IYJ$ excess and the Paschen and Brackett line fluxes. Although the kinematic structure of these lines points to an origin in the accretion flow and inner wind rather than the inner disk \citep{muz01,fol01,kwa10}, the unexpected finding of \citet{edw06} that the structure of the Pa$\gamma$ line profile depends on the 1 \micron\ veiling, with the broadest lines in stars with the largest veiling, makes it clear that the origin of these lines is not yet fully understood.  Thus, a contribution from the inner disk cannot be ruled out. More broadly, the complexity of the rich emission line spectrum of CTTS in the optical and near infrared is not fully explored and may yet reveal lines that are primarily formed in the inner gaseous disk \citep{ber98}.

Ongoing efforts to definitively identify the physical properties of the inner region of accretion disks should yield an enriched understanding of accretion in CTTS. The $IYJ$ excess is of particular interest in the CTTS that are the most heavily veiled, such as DR Tau, DG Tau A, RW Aur A, and AS 353A. Such CTTS are borderline class I/II young stellar objects, with large infrared excesses, large accretion rates, and spatially resolved micro-jets, and their large $IYJ$ excesses can provide a view of the star-disk interaction region in a phase of protostellar evolution that is usually heavily obscured in the optical and near infrared.

\section{CONCLUSIONS}

We have characterized the excess continuum emission in accreting T Tauri stars between 0.48 and 2.4 \micron. At 1 \micron\ the magnitude of the excess ranges from 0.1 to 3 times the photospheric flux, and it scales with the excess at shorter and longer wavelengths. We interpret this excess as arising from a source of continuum emission in addition to small hot ($T\sim8000$ K) accretion spots on the stellar surface at the base of accretion columns and large rings of warm ($T\sim1400$ K) dust at the sublimation radius in the disk.  The broad, smooth distribution of the $IYJ$ excess suggests it comes from a third component with a temperature between 2200 K and 5000~K and roughly blackbody emission characteristics. We cannot pin down a unique combination of temperature and filling factor to account for the emission, but among stars with high $IYJ$ excess, the filling factor is comparable to the surface area of the star if the temperature is 5000~K or more than an order of magnitude larger if the temperature is 2500 K.  Possible sources of the emission include warm annuli surrounding hot accretion spots in the shock-heated photosphere, accreting gas in funnel flows, or disk gas inside the dust sublimation radius. The luminosity from this region is comparable to the accretion luminosity found from ultraviolet and blue excesses, suggesting that accretion luminosities  have been underestimated by a factor of about two. The implication for disk accretion rates is more complicated, but a comprehensive campaign to obtain simultaneous spectrophotometry from the ultraviolet to the near infrared for a large sample of CTTS would provide further insight. Whatever its source, this excess emission contains a large fraction of the accretion energy released in the vicinity of the star.

\vspace{-0.1in}

\acknowledgments

We are grateful to Scott Dahm for flexibility in the scheduling of his HIRES time and for acquiring our HIRES spectra. We acknowledge stimulating conversations with Nuria Calvet, Laura Ingleby, and Scott Gregory, and we thank the anonymous referee for helpful comments. This work was partially supported by NASA grant NNG506GE47G issued through the Office of Space Science. The authors reverently acknowledge the cultural significance of the Mauna Kea summit to the indigenous Hawaiian community.  We are most fortunate to have had the opportunity to conduct observations with IRTF and the Keck telescopes from this mountain.

\end{document}